 \documentclass[11pt]{JHEP3}

\usepackage{amssymb,amsfonts,amsmath}
\newcommand{\be}{\begin{equation}}
\newcommand{\ee}{\end{equation}}

\newcommand{\mn}{{\mu\nu}}

\newcommand{\co}{\mathcal{O}}

\newcommand{\bi}{\begin{itemize}}
\newcommand{\ei}{\end{itemize}}

\newcommand{\bnum}{\begin{enumerate}}
\newcommand{\enum}{\end{enumerate}}

\newcommand{\bvb}{\begin{verbatim}}

\newcommand{\ba}{\begin{eqnarray}}
\newcommand{\ea}{\end{eqnarray}}

\newcommand{\eq}[1]{(\ref{#1})}

\author{Xing Wu\\

{\it Department of Physics, North University of China, \\ 3 Xueyuan Road, Taiyuan, Shanxi 030051, China}\\

E-mail:   \email{xwu@nuc.edu.cn}
}

\abstract{Recent developments have revealed a new phenomenon, i.e. the residues of the poles of the holographic retarded two point functions of generic operators vanish at certain complex values of the frequency and momentum. This so-called pole-skipping phenomenon can be determined holographically by the near horizon dynamics of the bulk equations of the corresponding fields. In particular, the pole-skipping point in the upper half plane of complex frequency has been shown to be closed related to many-body chaos, while those in the lower half plane also places universal and nontrivial constraints on the two point functions. In this paper, we study the effect of higher curvature corrections, i.e. the stringy correction and Gauss-Bonnet correction, to the (lower half plane) pole-skipping phenomenon for generic scalar, vector, and metric perturbations. We find that at the pole-skipping points, the frequencies $\omega_n=-i2\pi nT$ are not explicitly influenced by both $R^2$ and $R^4$ corrections, while the momenta $k_n$ receive corresponding corrections. }

\title{Higher curvature corrections to pole-skipping}

\begin{document}

\maketitle
\tableofcontents

\newpage

\section{Introduction}

    In recent years, progress in quantum many-body chaos has attracted much interest. In particular, developments in the gauge/gravity duality \cite{Maldacena:1997re} have exhibited close connection between black hole physics and  chaos in quantum many-body systems. Usually, the chaotic behavior is characterized by the out-of-time correlation functions (OTOCs), from which two characteristic parameters can be obtained, i.e. the quantum Lyapunov exponent $\lambda_L$, and the butterfly velocity $v_B$ \cite{Larkin:1969aa,Shenker:2013pqa,Roberts:2014isa,Shenker:2014cwa}.  % A typical result characterizing chaos in certain thermal systems is given by the form
    Within the framework of holography, the OTOC can be obtained from the shock wave analysis in the dual gravity theory \cite{Shenker:2013pqa}. In particular, black holes are argued to be the fastest scramblers \cite{Sekino:2008he,Susskind:2011ap}, which saturate an upper bound on the Lyapunov exponent \cite{Maldacena:2015waa} $\lambda_{L}=2\pi T$.

    Later, it was argued that besides the OTOCs, which are essentially {\it four} point functions, quantum chaos can also be manifested  in the retarded {\it two} point functions. Numerical analysis in \cite{Grozdanov:2017ajz} first indicates that information of chaos previously obtained via holography from non-linear shock wave geometry can be extracted from hydrodynamic sound modes in linearized gravitational perturbation equation. More precisely, at certain imaginary values of frequency $\omega_*$ and momentum $k_*$ of the sound pole of the retarded stress-energy two point function, the residue of the pole also vanishes, i.e., the pole is ``skipped". At the pole-skipping point, the Lyapunov exponent can be read off from the imaginary frequency $\omega_*=i\lambda_L$, while the butterfly velocity can be determined from the dispersion relation right at the point $\omega_*=v_Bk_*$.  In  \cite{Blake:2017ris}, this pole-skipping phenomenon was also explained in terms of the shift symmetry of an effective hydrodynamic description. The pole-skipping was also analytically studied in \cite{Blake:2018leo} as an universal behavior near the horizon where the time-time component of the Einstein equation (in ingoing Eddington-Finkelstein coordinates) vanishes such that the dual retarded two point function is not uniquely defined.     The pole-skipping phenomenon has also been checked to hold for SYK system \cite{Gu:2016oyy} and 2D CFT with large central charge \cite{Haehl:2018izb}. The stringy correction and Gauss-Bonnet high curvature correction to pole-skipping was investigated in \cite{Grozdanov:2018kkt}, where imaginary frequency $\omega_*$ receives no explicit correction\footnote{In particular, the frequency is still given by $\omega_*=i2\pi T$, although $T$ implicitly involves stringy or Gauss-Bonnet correction.
     }
     while the butterfly velocity $v_B$ does receives correction, which was shown to agree with the results obtained using the shock wave solution as in \cite{Roberts:2014isa}. Pole-skipping for CFT in hyperbolic space dual to AdS-Rindler geometry is also discussed in \cite{Ahn:2019rnq,Haehl:2019eae}.

 %    The pole-skipping phenomenon has been checked to hold for SYK system [1609.07832] and 2D CFT with large central charge [1808.02898]. In [1809.01169], pole-skipping is holographically related to the near horizon behavior of the bulk Einstein equation. More precisely, the time-time component of the bulk equation vanishes precisely at the pole-skipping point.

    Recently, the near horizon analysis is generalized in \cite{Blake:2019otz} to equations of bulk fields dual to spin-0, spin-1 and spin-2 operators, and pole-skipping is found to exist in retarded two point functions of these operators. However, these pole-skipping points appear in the {\it lower half} plane of the complex frequency, in contrast to the aforementioned pole-skipping point of chaos located in the {\it upper half} plane at $\omega=+i2\pi T$. This indicates that pole-skipping may not always be directly related to quantum chaos, but could be a consequence of  a more general feature of near horizon bulk equations. Relevant discussions can also be found in \cite{Natsuume:2019sfp,Natsuume:2019xcy,Grozdanov:2019uhi}.

    In this paper, we continue to study pole-skipping along the line of \cite{Blake:2019otz} by involving the stringy correction and Gauss-Bonnet correction. It turns out that the dependence of the frequencies at the pole-skipping points remain the same as in the uncorrected case, while the momenta receives corrections. This pattern agrees with that found in \cite{Grozdanov:2018kkt} for pole-skipping in the upper plane. Moreover, the upper half plane pole-skipping point can also be recovered and is shown to agree with the results obtained in \cite{Grozdanov:2018kkt}.

    This paper is organized as follows. Section \ref{section review} reviews the key ideas relevant to pole-skipping in the uncorrected background. In Section \ref{section stringy correction}, we discuss pole-skipping in the presence of the stringy correction and obtain the corresponding imaginary values of $\omega$ and $k$ for typical scalar operator, current operator, and stress-energy tensor, respectively. Similar analysis involving the Gauss-Bonnet term will be presented in Section \ref{section GB correction}. We conclude with a summary and discussion in the final section.

\section{Review of key ideas}\label{section review}
To elucidate the key ideas of \cite{Blake:2019otz} as well as \cite{Blake:2018leo} that are relevant to our discussion of pole-skipping, we will first consider a minimally coupled scalar field $\varphi$ in the uncorrected background, i.e. AdS$_5$ black brane\footnote{In this paper, the AdS radius is always set to unity for convenience.},
\be
ds^2=-r^2f(r)dv^2+2dvdr+r^2(dx^2+dy^2+dz^2).
\ee
where $f(r)=1-r_0^4/r^4$ with the horizon location $r_0$. Note the metric has already been written in ingoing Eddington-Finkelstein coordinates. The scalar field obeys the equation of motion \footnote{One may well consider the equivalent form $\nabla_\mu\nabla^\mu\varphi-m^2\varphi=0$. Note in that case, the near horizon expansion of the perturbation equation would in general be different at each order due to the extra $\sqrt{-g}$ factor. Of course, the physics will remain the same. Here we simply follow the convention used in \cite{Blake:2019otz} for the sake of comparison.}(EOM)
\be\label{scalar eom}
\partial_\mu(\sqrt{-g}\partial^\mu\varphi)-\sqrt{-g}m^2\varphi=0,
\ee
Assuming that the perturbation depends only on $x$, in addition to time and the radial direction, using the Fourier transform $\varphi(v,r,x)\rightarrow e^{-i\omega v+ikx}\phi(r)$,  the EOM   becomes
\be\label{scalar eom fourier AdS}
r^5 f(r) \phi ''(r)+\left[r^5 f'(r)+5 r^4 f(r)-2 i r^3 \omega \right]\phi '(r)  + \left(-k^2 r-m^2 r^3-3 i r^2 \omega \right)\phi (r)=0,
\ee
where a prime indicates taking derivative with respect to $r$.

The holographical dual of  the bulk scalar field is a scalar operator of dimension $\Delta$ determined by the mass of the bulk field via $\Delta(\Delta-4)=m^2$. The retarded two point function of the scalar operator (in Fourier space) is given as \cite{Son:2002sd,Herzog:2002pc}
\be\label{retarded G}
G^{R}(\omega,k)\sim \frac{B(\omega,k)}{A(\omega,k)},
\ee
where $A$ and $B$ are coefficients in the asymptotic expansion of the scalar field near the boundary
\be
\phi\rightarrow A r^{\Delta-4}+B r^{-\Delta}.
\ee
Moreover, the field obeys the ingoing wave condition at the horizon. Then the poles of $G^R$, i.e. $A(\omega,k)=0$, just defines the quasi-normal mode spectrum \cite{Son:2002sd,Kovtun:2005ev} for the scalar field perturbation.

As argued in \cite{Blake:2019otz}, there exists certain values $(\omega_n,k_n)$, referred to as pole-skipping points, at which both $A$ and $B$ vanish, such that the retarded two point function is not well defined. As first indicated in \cite{Blake:2018leo} and further explored in \cite{Blake:2019otz}, the pole-skipping points manifest themselves in the dual gravity theory as some special locations in $\omega$ and $k$ for the bulk EOM. This can be understood as follows. Since we are in ingoing Eddington-Finkelstein coordinates where the metric functions are regular near the horizon $r_0$, we can insert the near horizon expansion of the scalar field
\be\label{scalar NH expansion}
\phi(r)=\sum_{n=1}\phi_{n-1}(r-r_0)^{n-1}
\ee
into EOM \eq{scalar eom fourier AdS}, and then expand the EOM near the horizon $r_0$. Then a (infinite) series of perturbed EOM in the order of $(r-r_0)$ can be obtained as
\ba
\co[(r-r_0)^0]:\ \ \  0&=&C_{00}\phi_0+C_{01}\phi_1,\nonumber \\ % \label{review: order 1}\\
\co[(r-r_0)^1]:\ \ \   0&=&C_{10}\phi_0+C_{11}\phi_1+ C_{12}\phi_2, \nonumber \\ %\label{review: order 2}\\
%\co[(r-r_0)^2]:\ \ \   0&=&C_{20}\phi_0+C_{21}\phi_1+C_{22}\phi_2+ C_{23}\phi_3,\\
&\vdots&\nonumber\\
\co[(r-r_0)^{n-1}]:\ \ \   0&=&C_{n-1~0}~\phi_0+C_{n-1~1}~\phi_1+\dots+C_{n-1~n-1}~\phi_{n-1}+ C_{n-1~n}~\phi_{n},\nonumber\\
&\vdots& \label{review: order n}
\ea
where the coefficients $C_{ij}$ are functions of $\omega$ and $k$. For generic $\omega$ and $k$,   one can solve for $\phi_1$ in terms of $\phi_0$ from the $\co[(r-r_0)^0]$ equation in \eq{review: order n}, and iteratively obtain other $\phi_i$ in terms of $\phi_0$ order by order. Then the ingoing solution is uniquely determined (up to the normalization associated with $\phi_0$), and the retarded two point function \eq{retarded G} is well defined.

However, when $C_{01}=0$, which gives $\omega= \omega_1\equiv-i2\pi T$, $\phi_1$ cannot be determined by $\phi_0$. Moreover, when $C_{00}$ is also vanishing, leading to certain value $k=k_1$, $\phi_0$ is also unconstrained. This gives the first pole-skipping points $(\omega_1,k_1)$. Now the two free parameters $\phi_0$ and $\phi_1$ imply that the ingoing solution is not uniquely defined, leading to the pole-skipping phenomenon in the two point function in the dual field theory.

Similarly, other pole-skipping points with higher frequencies can be obtained. Indeed, in the $\co[(r-r_0)^{n-1}]$ equation of \eq{review: order n},
the vanishing of the coefficient $C_{n-1~n}$ gives $\omega=\omega_{n}\equiv-i2\pi Tn$, and thus implies that $\phi_{n}$ is unconstrained. Moreover, with $C_{n-1~n}=0$ and generic values for $k$, the first $n$ equations as a set of algebraic equations for the first $n$ variables $(\phi_0,\dots,\phi_{n-1})$ imply that all of these variables should vanish, unless the momentum $k$ takes some special values $k_{n}$ arising from the vanishing of the determinant of the  coefficient matrix
\be\label{C matrix}
M_{n}\equiv\begin{pmatrix} C_{00}&C_{01}&0&\dots \\ C_{10}&C_{11}& C_{12}& 0 &\dots \\ \dots \\ C_{n-1~0} &C_{n-1~1} &C_{n-1~2} &C_{n-1~3}&\dots  &C_{n-1~n-1}
\end{pmatrix}.
\ee
Note, in particular,  {$M_{1}=C_{00}$}.
In sum, the two conditions,
\be\label{two conditions}
C_{n-1~n}=0,\ \ \ \det M_n=0,
\ee
 together determine the locations of the pole-skipping points $(\omega_{n},k_{n})$. Note that the algebraic equation $\det M_n=0$ in general produces $n$ complex values for $k_{n}$.

In the following, we will investigate the effect of the stringy correction  and Gauss-Bonnet correction to the pole-skipping phenomenon by performing the near horizon analysis as above. In particular, we will work out similar equations as \eq{review: order n} for various types of bulk fields, from which the pole-skipping points of the corresponding retarded two point functions are determined by the two conditions in \eq{two conditions}.

\section{Stringy correction to pole-skipping}
\label{section stringy correction}
\subsection{Setup}

The finite 't Hooft coupling correction in the $SU(N_c)$ $\mathcal{N}=4$   supersymmetric Yang-Mills theory (SYM) in the large $N_c$  limit is holographically dual to the stringy $\alpha'$ correction in supergravity. More precisely, the leading finite $\lambda$ correction comes at $\co(\lambda^{-3/2})$, corresponding to the $\alpha'^3$ correction to Einstein gravity. The usual starting point for discussing the stringy correction (e.g. in \cite{Gubser:1998nz,Pawelczyk:1998pb}) is the 10D type IIB low-energy effective action \cite{Grisaru:1986vi,Freeman:1986zh,Park:1987jp,Gross:1986iv}
\be
S_{10}=\frac{1}{2\kappa_{10}^2}\int d^{10}x\sqrt{-g}\left[ R^{(10)}-\frac{1}{2}(\partial\Phi)^2-\frac{1}{4\cdot 5!}(F_5)^2+\dots+\gamma e^{-\frac{3}{2}\Phi}W^{(10)}+\dots\right],
\ee
where  $\kappa_{10}^2$ is essentially the 10D gravitational constant, $R^{(10)}$ is the 10D Ricci scalar, $\gamma=\alpha'^{3}\zeta(3)/8\sim\lambda^{-3/2}$ is the parameter for the leading order $\alpha'$ correction, $W^{(10)}$ is a fourth order high curvature term, which can be expressed in terms of the Weyl tensor $C_{\mn\alpha\beta}$ as
\be\label{W expression}
W^{(10)}=C^{\mn\rho\sigma}C_{\alpha\nu\rho\beta}C_\mu^{~\gamma\delta\alpha}C^\beta_{~\gamma\delta\sigma}
+\frac{1}{2}C^{\mu\sigma\nu\rho}C_{\alpha\beta\nu\rho}C_\mu^{~\gamma\delta\alpha}C^\beta_{~\gamma\delta\sigma}.
\ee
Since the dilaton $\Phi$ decouples from the gravitational perturbation equation to leading order in the $\alpha'$ correction, it can be simply neglected in the following. As argued in \cite{Myers:2008yi}, the RR 5-form $F_5$ and other fields are also irrelevant for our purpose. We will follow \cite{Buchel:2008ae} (see also \cite{Grozdanov:2018kkt}) and only consider the dimensionally reduced 5D action with a correction term
\be\label{Stringy action}
S=\frac{1}{2\kappa_5^2}\int d^5x\sqrt{-g}(R+12+\gamma W),
\ee
where $\kappa_5$ gives the effective 5D gravitational constant, $W$ is just given by \eq{W expression} with the 10D Weyl tensors replaced by the  5D ones. To our purpose in this paper, we will focus on the 5D action \eq{Stringy action} and study the effect of the leading order $\gamma$ correction on the pole-skipping phenomenon.

%The dots represent other fields which are irrelevant for our discussion here.

The background solution is the $\gamma$-corrected black brane  \cite{Gubser:1998nz,Pawelczyk:1998pb}
\be
ds^2=r^2\left[-f(r)Z_tdt^2+dx^2+dy^2+dz^2\right]+Z_r\frac{dr^2}{r^2f},
\ee
where $f(r)=1-r_0^4/r^4$,   and
\ba
Z_t&=&1-15\gamma\left(5\frac{r_0^4}{r^4}+5\frac{r_0^8}{r^8}-3\frac{r_0^{12}}{r^{12}}\right),\\
Z_r&=&1+15\gamma\left(5\frac{r_0^4}{r^4}+5\frac{r_0^8}{r^8}-19\frac{r_0^{12}}{r^{12}}\right).
\ea
The Hawking temperature receives the leading order correction
\be\label{stringy T}
T=T_0(1+15\gamma ),
\ee
with $T_0=r_0/\pi$ being the uncorrected temperature.  To facilitate our near horizon analysis, we change to  ingoing Eddington-Finkelstein coordinates
\be
v=t+r_*,\ \ \ dr_*=\frac{dr}{r^2f(r)}\sqrt{\frac{Z_r}{Z_t}}.
\ee
Then, the metric takes the form
\be\label{Stringy metric}
ds^2=r^2\left[-f(r) Z_{vv}dv^2+dx^2+dy^2+dz^2\right]+2Z_{vr}dvdr,
\ee
where $Z_{vv}=Z_t$, and
\be
Z_{vr}=\sqrt{Z_tZ_r}=1-120\gamma\frac{r_0^{12}}{r^{12}},
\ee
up to $\co(\gamma^2)$ terms which are dropped.

\subsection{Scalar field}

Let us begin by considering pole-skipping in the case of a generic scalar operator. In the $\mathcal{N}=4$ SYM theory with leading finite 't Hooft coupling correction at $\co(\lambda^{-3/2})$, a scalar operator  is dual to a bulk scalar field in the above background \eq{Stringy metric}. As shown in \cite{Blake:2019otz}, the retarded two point function exhibits pole-skipping at frequencies $\omega_n=-i2\pi Tn$ with $n=1,2,3,\dots$, and corresponding complex momenta $k_n$. Here we further explore this phenomenon by performing the near horizon analysis of the scalar EOM in the presence of the stringy correction.

Compared to the uncorrected case, the EOM in the background \eq{Stringy metric}  receives a $\gamma$-dependent source term, and equation \eq{scalar eom fourier AdS} receives a $\gamma$-dependent source term as
\be\label{scalar eom fourier}
r^5 f(r) \phi ''(r)+\left[r^5 f'(r)+5 r^4 f(r)-2 i r^3 \omega \right]\phi '(r)  + \left(-k^2 r-m^2 r^3-3 i r^2 \omega \right)\phi (r)=\gamma S_1,
\ee %1904.12883StringyCorrectionScalar.nb
where   the source $S_1$ is given in  Appendix \ref{app:stringy:scalar}. Inserting the near horizon expansion \eq{scalar NH expansion}, the above EOM \eq{scalar eom fourier} leads to a series of equations of the form \eq{review: order n}.
The first few coefficients $C_{ij}$ are listed in Appendix \ref{app:stringy:scalar}. In particular, the coefficients in the leading $\co[(r-r_0)^0]$ equation become
\ba
C_{00}&=&-  k^2-{r_0} \left(m^2  {r_0}+3 i \omega \right)+\gamma  120   \left(k^2+m^2  r_0^2\right),\\
C_{01}&=&\left( {r_0}^4 f'_0-2 i r_0^2 \omega \right)+\gamma  15  r_0^4 f'_0,\label{M00}
\ea
where $f'_0$ denotes $f'(r_0)$.
%For generic $\omega$ and $k$, this relation determines $\phi_1$ in terms of $\phi_0$. Performing this near horizon analysis order by order in $(r-r_0)$, one can determine  higher order coefficients $\phi_n$ in terms of the value at horizon $\phi_0$, and correspondingly the two point function on the dual field theory is uniquely defined. For specific complex values of $\omega$ and $k$, however, this may not hold.

The two conditions \eq{two conditions}, i.e. $C_{00}=0$ and $C_{01}=0$ in the present case, can be used to find the correction to the first pole-skipping point. Inserting the temperature \eq{stringy T}, one can easily see that the coefficient of $\phi_1$ vanishes at the frequency
\be
\omega_1=-2\pi Ti.
\ee
It should be emphasized that $T$ is the $\gamma$-corrected temperature in \eq{stringy T}. At the same time, the coefficient of $\phi_0$ vanishes at
\be\label{stringy scalar k1}
k^2_1=-(m^2 +6) r_0^2-\gamma 810  r_0^2= -(m^2+6) \pi ^2T^2+\gamma (30 m^2-630) \pi ^2 T^2,
\ee
which can be written in terms of field theory quantities as
\be
k^2_1= -[\Delta(\Delta-4)+6] \pi ^2T^2+\gamma [30 \Delta(\Delta-4)-630] \pi ^2 T^2.
\ee
Keeping in mind that $\gamma$ is perturbative, one can see that $k_1$ takes the imaginary value  $k_1= i[r_0(m^2+6)^{1/2}+\gamma 405 r_0(m^2+6)^{-1/2}]$. These values are shifted compared to the result in equation (2.16) of \cite{Blake:2019otz}, due to the stringy correction. Moreover, analysis of $\co[(r-r_0)^{n-1}]$ equation  indicates $C_{n-1~n}\propto [2\pi Tn-i\omega]$, the same as the uncorrected result, while the momenta $k_n$ receive explicit $\gamma$ corrections.
For example, $k_2$ and $k_3$ are given from
\ba
0&=&r_0^2 \left[k_2^4+2 k_2^2(m^2+12) r_0^2+(m^4+16 m^2+96) r_0^4\right]\nonumber\\
&&-120\gamma \left[2 k_2^4 r_0^2+k_2^2 \left(4 m^2-3\right) r_0^4+2 \left(m^4-5 m^2-444\right) r_0^6\right],\\
0&=&-r_0^3 \left[k_3^6+3 k_3^4 (m^2+8) r_0^2+k_3^2 (3 m^4+40 m^2+96) r_0^4+m^2 (m^4+16 m^2+96) r_0^6\right]\nonumber\\
&&+360 \gamma  r_0^3 \left[k_3^6+k_3^4 (3 m^2-65) r_0^2+3 k_3^2 (m^4-45 m^2-664) r_0^4\right.\nonumber\\
&&\left.+(m^6-70 m^4-1416 m^2-6912) r_0^6\right].
\ea
 {
In particular, a compact expression, perturbative in $\gamma$, for $k_2^2$ can be solved  as
\be
k_2^2=-(12+m^2\pm2 \sqrt{2} \sqrt{m^2+6})r_0^2 \pm\gamma\frac{45 \sqrt{2}r_0^2 \left(156-3 m^2\mp34 \sqrt{2} \sqrt{m^2+6}\right)}{\sqrt{m^2+6}},
\ee
where the first term recovers the result of \cite{Blake:2019otz} in the absence of the stringy correction, and the second term is the $\gamma$-correction to $k_2^2$. The analytic expression for $k_3^2$ is too lengthy to be listed here, and higher $k_n^2$ in general must be solved numerically. Thus, in the following, except for the case of vector perturbations with stringy corrections \eq{stringy vector k3}, only compact expressions for $k_1^2$ and $k_2^2$ will be presented.
%Or, maybe not: the three solutions for $k_3^2$ are
%\ba
%k_3^2&=&-m^2r_0^2-\gamma 25920 r_0^2,\\
%k_3^2&=&-(m^2 +12 \pm2 \sqrt{2} \sqrt{m^2 +6})r_0^2 -\gamma45 r_0^2\frac{ 3 \sqrt{2} (m^2+16) \sqrt{(m^2+6) }\pm(86 m^2-528) }{3 \sqrt{2} \sqrt{\left(m^2+6\right) }\pm(m^2+6) }
%\ea
}

In sum, the near horizon analysis reveals the pole-skipping points at $\omega_n=-2\pi n Ti$  and the corresponding complex $k_n$, for generic scalar operators. Compared with the uncorrected result in \cite{Blake:2019otz}, although the temperature dependence of the frequencies remains the same,  the relations between $\omega_n$ and $k_n$  receive   $\co(\gamma)$ corrections.  This is similar to the result in \cite{Grozdanov:2018kkt} for the pole-skipping point in the upper half plane of complex frequency. There, the modification to $k_*$ leads to $\gamma$-corrected butterfly velocity $v_B$. Note that in our case here, $\omega_n/k_n$ at the pole-skipping point is in general not directly related to $v_B$.

\subsection{Vector field}

Consider a $U(1)$ current operator $J^\mu$, dual to a Maxwell vector field $A_\mu$ in the background \eq{Stringy metric}, described by the EOM
\be\label{vector eom}
\partial_\mu(\sqrt{-g}Z(\Phi)F^\mn)=0,
\ee
where $\Phi$ controls the effective coupling of the gauge field\footnote{Following \cite{Blake:2019otz}, the effective Maxwell coupling $Z(\Phi)$ is introduced to make the discussion as general as possible. But for convenience, we will assume $\Phi=\Phi(r)$ and thus $Z$ is essentially only a function of $r$.
}.
In the spirit of \cite{Policastro:2002se,Kovtun:2005ev}, the vector perturbations can be classified according to the $O(2)$ symmetry in the plane normal to the direction of the momentum, chosen to be the $x$ direction here.
The perturbations $A_y$ and $A_z$ as $O(2)$ vectors are in the transverse channel, whereas $A_v$, $A_r$ and $A_x$ as $O(2)$ scalars are in the longitudinal (or, diffusive) channel. EOMs of perturbations in different channels decouple. Since EOMs in the transverse channel are two decoupled equations similar to that of the above minimally coupled scalar field, the analysis and results in this channel are similar to the above results. So we will not discuss this channel in detail, and only focus on the longitudinal channel, where there is a hydrodynamic diffusion mode. We will also use the radial gauge $A_r=0$.

In the longitudinal channel, the perturbations are coupled to each other.  However, $A_v$ and $A_x$ can form a gauge invariant variable, i.e. the electric field $E$, defined by
\be\label{vector master field}
E=\omega A_x+kA_v.
\ee
Then the three equations for $A_v$ and $A_x$ can be combined into a single equation for $E$,
\be\label{stringy vector EOM}
E''+A_{E} E'+B_{E} E=0,
\ee
where the coefficients $A_{E}$ and $B_{E}$ are given in Appendix \ref{app:stringy:vector}.
Note that the two coefficients depend on $\gamma$, and will be expanded to $\co(\gamma)$ in the following calculation.
%Inserting the expressions for $Z_{vv}$ and $Z_{vr}$, and keeping only $\co(\gamma)$ corrections,

To perform the near horizon analysis, one can insert the expansion
\be
E=\sum_{n=1}E_{n-1}(r-r_0)^{n-1}
\ee
into \eq{stringy vector EOM} and expand it near the horizon. Analyzing each order in $(r-r_0)$, one can obtain a set of equations of the same form as \eq{review: order n}. In particular, applying the conditions \eq{two conditions}, the leading order equation gives the first pole-skipping points at
\ba
\omega_1&=&-i2\pi T,\\
 k^2_1&=&2r_0^2\left(1+r_0\frac{Z'_0}{Z_0}\right)\left(1+135\gamma\right)=2\pi^2 T^2\left[\left(1+\pi T\frac{Z'_0}{Z_0}\right)+\gamma\left(105+90\pi T \frac{Z'_0}{Z_0}\right)\right].\label{stringy vector k1}
\ea

Again, the stringy effect only produces a $\gamma$-correction to $T$, but the $T$-dependence of $\omega_1$ is not modified. However,  $k_1$ receives an explicit $\gamma$-correction.
  {To focus on the effect of the stringy correction, we  take $Z=1$ for simplicity. Then
\be\label{stringy vector k1b}
k_1^2=2r_0^2(1+135\gamma),
\ee
and the momenta corresponding to $\omega_2$ and $\omega_3$ are determined from}
\ba
0&=&k_2^4+8 k_2^2 r_0^2-32 r_0^4-\gamma 60 \left(5 k_2^4-26 k_2^2 r_0^2+560 r_0^4\right),\nonumber\\
0&=&k_3^6+42 k_3^4 r_0^2+300 k_3^2 r_0^4-1800 r_0^6,\nonumber\\
&&-\gamma  90  \left(5 k_3^6+19 k_3^4 r_0^2-4904 k_3^2 r_0^4+90492 r_0^6\right).
\ea
 {
The expressions for $k_2^2$ can be solved as
\ba
k_2^2&=&-4 \left(1\pm\sqrt{3}\right) r_0^2-\gamma 60  \left(33\pm41\sqrt{3}\right) r_0^2.\label{stringy vector k2}
\ea
In this case, the three solutions for $k_3^2$ also take compact forms
\ba
k_3^2&=&-30r_0^2+\gamma 22446  r_0^2,\nonumber \\
k_3^2&=&-2(3\pm2 \sqrt{6})r_0^2-\gamma\frac{30 (257 \sqrt{6}\pm1761) }{\sqrt{6}\mp1}r_0^2.\label{stringy vector k3}
\ea
%One can easily check that the corresponding $\co(\gamma)$ terms are still positive. Thus, the first three $k_n$ all   have real solutions, even in the presence of the $\gamma$-corrections. It is expected that this continues to be valid for higher $k_n$, despite the lack of a rigorous proof.
}

It is easy to see from \eq{stringy vector k1b} that $k_1^2$ is positive, or, $k_1$ is real, in contrast to the case of scalar operator \eq{stringy scalar k1} where $k_1$ is imaginary. Moreover, the $\co(\gamma^0)$ solutions $k_2^2=-4(1-\sqrt{3})r_0^2$ and $k_3^2=-2(3-2\sqrt{6})r_0^2$ are positive, corresponding to real $k_2$ and $k_3$. Since the $\gamma$-corrections are perturbations, which should not change the sign of  the leading order $k_n^2$, these solutions for $k_2$ and $k_3$ are real in the presence of the stringy correction. In general, it is expected that $k_n$ has $n$ values, of which at least one is real.

The real values of $k_n$ are related to the diffusion mode in this channel. It is well-known that \cite{Policastro:2002se,Kovtun:2005ev} in the hydrodynamic limit $\omega\ll T$ and $k\ll T$,  the diffusion mode has a pole in the retarded two point function, $\omega=-iD_R k^2$ with $D_R$ the R-charge diffusion constant, which receives the string correction, c.f. \cite{Benincasa:2005qc}. As argued in \cite{Grozdanov:2017ajz,Blake:2018leo,Blake:2019otz}, the pole-skipping phenomenon places nontrivial constraints on the dispersion relation $\omega(k)$ at $|\omega|\sim T$, beyond the hydrodynamic region.  {In other words, the dispersion relation $\omega(k)$ of the hydrodynamic diffusion mode approaches $(\omega,k)=(0,0)$ in the form of the diffusion pole, and passes through the pole-skipping points $(\omega_n,k_n)$ for $k$ large relative to $T$. }

 {
By comparing the  magnitude of the numerical coefficient of the   $\co(\gamma)$ correction  relative to that of the leading $\co(\gamma^0)$ term  in the expressions for $k_1^2$, $k_2^2$ and $k_3^2$ in \eq{stringy vector k1b}, \eq{stringy vector k2} and \eq{stringy vector k3}, one can see that the ratio becomes larger for higher $k_n^2$. Indeed, for $k_1^2$, the ratio is $135$  in \eq{stringy vector k1b}. For $k_2^2$, the largest ratio in the two solutions in \eq{stringy vector k2} is $|60(33-41\sqrt{3})|/|4(1-\sqrt{3})|\approx779$. For $k_3^2$, the largest ratio in the three solutions in \eq{stringy vector k3} is approximately 3132. Recall that these results are all obtained with $\gamma$ treated as a perturbative parameter. So,  for the $\co(\gamma)$ terms to be legitimate perturbations, $\gamma$
should be constrained by an upper bound $\gamma_1\equiv 1/135$ for $k_1$, $\gamma_2\equiv 1/779$ for $k_2$, and $\gamma_3\equiv 1/3132$, with tighter bounds for higher $k_n$ being expected.\footnote{This is also important for numerical studies. For example, if one takes $\gamma=0.001$, one would only find real solution for $k_1$, but not for $k_2$ and $k_3$, because this $\gamma$ is smaller than $\gamma_1$ for $k_1^2$, but larger than the bounds $\gamma_2$ and $\gamma_3$, for $k_2$ and $k_3$.}
In other words, higher $k_n^2$ becomes more sensitive to $\gamma$-corrections.\footnote{Note that this sensitivity to $\gamma$ is essentially also present in all other cases, including scalar field perturbation and metric perturbations. See, e.g., \eq{stringy shear k1 sol} and \eq{stringy shear k2 sol}.  So, the discussion for the typical results here will not be repeated in other sections.}
Similar issue was also discussed in the study of the finite coupling corrections to quasinormal modes \cite{Waeber:2015oka,Buchel:2018eax}, where the upper bound on $\gamma$ for the quasinormal modes is significantly increased by an effective resummation of a subset of higher order corrections arising solely from the first order $\co(\gamma)$ correction. We will not pursue a possible resummation scheme here, but leave it for future work.   }

\subsection{Metric perturbation}
\label{sec:stringy:metric perturbation}
In order to study pole-skipping of the retarded two point function of energy momentum tensor, we consider metric perturbations to the background \eq{Stringy metric}, $g_\mn+h_\mn$. We focus on the Fourier transform $h_\mn(v,r,x)\rightarrow e^{-i\omega v+ikx}h_\mn(r)$. For simplicity, we assume the radial gauge $h_{r\mu}=0$. Then the perturbations can be classified by the $O(2)$ symmetry along the $yz$ plane into three decoupled channels:
\bi
    \item
    $O(2)$ tensor, scalar channel: $h_{yz}$;
    \item
    $O(2)$ vector, shear channel: $h_{v\alpha}$ and $h_{x\alpha}$, $\alpha=y,z$; % or $h_{vz}$ and $h_{xz}$
    \item
    $O(2)$ scalar, sound channel: $h_{vv}$, $h_{vx}$, $h_{xx}$, $h_{aa}\equiv h_{yy}+h_{zz}$.
\ei
%In the scalar channel, the gauge invariant variable $h_{y}^z=h_{yz}/r^2$ obeys the same equation as the minimally coupled scalar field discussed above. So similar results hold here for the near horizon analysis of $h_y^z$ and for the pole-skipping property of $G^R_{T^y_zT^y_z}$. We will not repeat the details.

 {
In Einstein gravity,  the gauge invariant variable $h_{y}^z=h_{yz}/r^2$ in the scalar channel obeys the same equation as a minimally coupled massless scalar field in the same background geometry. In the presence of higher curvature corrections,  the EOM of $h_y^z$ is not exactly the same as that of the scalar field. However, the qualitative features of the pole-skipping results are not significantly different from that of the scalar field. Moreover, there is no hydrodynamic mode in this channel \cite{Kovtun:2005ev,Grozdanov:2016vgg}. Therefore, in this paper, we will not present the detailed results in this channel, and only focus on the shear and sound channels where there are interesting hydrodynamic modes.
}

\subsubsection{Shear channel}
In the shear channel,  we consider the metric perturbations with only $h_{vy}$ and $h_{xy}$ non-vanishing. To obtain the linearized equations in the presence of the stringy correction, following \cite{Buchel:2004di,Benincasa:2005qc}, it is more convenient to insert the metric ansatz into the action \eq{Stringy action}, which is then expanded to quadratic order in $h_\mn$ to give an effective action for the perturbations, from which the linearized equations for $h_\mn$ follow.\footnote{
Of course, in general one should only insert the metric ansatz into the equation of motion, not the action. Here this is justified by the particular symmetries in the problem. Besides, the gauge condition $h_{r\mu}=0$ should also only be imposed on the level of the equation of motion. We must keep $h_{r\mu}\neq 0$ in the action in order to obtain the complete equations.
}
The two perturbations can be combined into one gauge invariant  variable, also referred to as ``master field",
\be\label{stringy:shear:Z1}
Z_1=\frac{1}{r^2}(\omega h_{xy}+k h_{vy}),
\ee
which obeys a single second order differential equation
\be\label{Z eq form}
Z_1''+AZ_1'+BZ_1=\gamma (M_0Z_1+M_1Z_1'),
\ee
where the coefficients $A$, $B$, $M_0$ and $M_1$ are given in Appendix \ref{app:stringy:shear}. Its derivation is rather tedious, and a schematic strategy of derivation is given in Appendix \ref{app:derivation:master equation}.

The near horizon analysis by inserting
\be
Z_1=\sum_{n=1} Z_{1n-1}(r-r_0)^{n-1}
\ee
into \eq{Z eq form} and expanding in $(r-r_0)$ leads to a series of equations of the same form as \eq{review: order n}, with $\phi_i$ replaced by $Z_{1i}$. The conditions \eq{two conditions} give again $\omega_{n}=-i2\pi T n$ with $k_{n}$ receiving explicit $\gamma$-corrections. The first three $k_{n}^2$ are determined by
\ba
0&=&k_1^2-6 r_0^2+\gamma  \left(-\frac{48 k_1^4}{r_0^2}+47 k_1^2+5868 r_0^2\right),\nonumber\\
0&=&k_2^4-96 r_0^4+\gamma  \left(-\frac{96 k_2^6}{r_0^2}-1826 k_2^4+19200 k_2^2 r_0^2+844416 r_0^4\right),\nonumber\\
0&=&k_3^6+30 k_3^4 r_0^2-180 k_3^2 r_0^4-4824 r_0^6\nonumber\\
&&+3 \gamma  \left(-\frac{48 k_3^8}{r_0^2}-3473 k_3^6-57160 k_3^4 r_0^2+1662060 k_3^2 r_0^4+61050672 r_0^6\right).
\ea
 {
Compact expressions for $k_1^2$ and $k_2^2$ are
\ba
k_1^2&=&6 r_0^2-\gamma 4422 r_0^2,\label{stringy shear k1 sol}\\
k_2^2&=&\pm4 \sqrt{6} r_0^2-\gamma 4r_0^2 (1248\pm3485\sqrt{6}  ).\label{stringy shear k2 sol}
\ea
}

As in the longitudinal channel of vector perturbations, here the real solutions for $k_n$ correspond to nontrivial constraints of pole-skipping on the momentum diffusion mode beyond the hydrodynamic range.  {Unlike the case of vector perturbations, however, here the $\gamma$-corrections can cause the originally positive $\co(\gamma^0)$ solutions $k_1^2=6r_0^2$ and $k_2^2=4\sqrt{6}r_0^2$ to become negative, unless the parameter $\gamma<6/4422\approx0.0014$ for real $k_1$, and $\gamma<\sqrt{6}/(1248+3485\sqrt{6})\approx0.00025$ for real $k_2$. However, these are also the conditions for the $\co(\gamma)$ terms to be legitimate perturbations. Therefore, as long as $\gamma$ is treated as a  perturbative parameter, $k_n$ always have real solutions which recover the hydrodynamic dispersion relation at small $k$.
}
%This seems to indicate that, if one insists that pole-skipping at $(\omega_n,k_n)$ should constrain the hydrodynamic mode, then there should be an upper bound on $\gamma$ to ensure the $O(\gamma^0)$ positive solution $k_n^2$ remain positive with $\gamma$-correction added. For larger $\gamma$ beyond the bound, curvature corrections higher than $R^4$ in $W$ of \eq{Stringy action} should be taken into account.}

\subsubsection{Sound channel}

In the sound channel, following \cite{Buchel:2004di,Benincasa:2005qc} again, the relevant perturbations  can also be combined into one single master field
\be\label{Z sound}
Z_2=\frac{1}{r^2}\left[2k^2h_{vv}+4k\omega h_{vx}+2\omega^2h_{xx}-(\omega^2-k^2\alpha_{12})h_{aa}\right],
\ee
where
\be
\alpha_{12}=1+\frac{r_0^4}{r^4}+15 \gamma \frac{r_0^4}{r^{4}}\left(5 -40 \frac{r_0^8}{r^8}+21 \frac{r_0^{12}}{r^{12}}\right).
\ee
The equation for $Z_2$ also takes the same form \eq{Z eq form}, with the coefficients given in Appendix \ref{app:stringy:sound}.

The near horizon analysis again leads to a series of equations. Again, we have $\omega_n=-i2\pi Tn$, and $k_n$ receive $\gamma$-corrections. In particular, the first three $k_n^2$  are determined by  equations arising from $\det M_n=0$
\ba
0&=&k_1^4-4k_1^2r_0^2+36k_0^4-3\gamma\frac{   16 k_1^8+93 k_1^6 r_0^2+1472 k_1^4 r_0^4+21412 k_1^2 r_0^6+70416 r_0^8 }{r_0^2 \left(k_1^2+6 r_0^2\right)},\nonumber\\
0&=&k_2^4-8 k_2^2 r_0^2+96 r_0^4-2 \gamma\frac{ 48 k_2^8+1623 k_2^6 r_0^2+8060 k_2^4 r_0^4+370272 k_2^2 r_0^6+10132992 r_0^8}{r_0^2 \left(k_2^2+24 r_0^2\right)},\nonumber\\
0&=&k_3^6+18 k_3^4 r_0^2-148 k_3^2 r_0^4+4824 r_0^6\nonumber\\
&&-\gamma \frac{ 144 k_3^{10}+16293 k_3^8 r_0^2+552046 k_3^6 r_0^4+6126820 k_3^4 r_0^6+237060648 k_3^2 r_0^8+9890208864 r_0^{10}}{r_0^2(k_3^2 +54 r_0^2)}.\nonumber\\
\ea
 {
Compact expressions for $k_1^2$ and $k_2^2$ can be solved as
\ba
k_1^2&=&2 (1\pm 2 i \sqrt{2})r_0^2+\gamma 6 (301\mp 382 i \sqrt{2})r_0^2,\\
k_2^2&=&4 (1\pm i \sqrt{5})r_0^2-\gamma4(73 \pm 4273 i \sqrt{5})r_0^2.
\ea
}

The sound channel includes the metric perturbation $h_{vv}$, which is dual to energy $T^{00}$ in the field theory. In contrast to the above pole-skipping points at the lower half plane of complex $\omega$, the energy retarded two point function exhibits pole-skipping at the upper half plane $\omega_*=+i2\pi T$, as was originally studied in \cite{Grozdanov:2017ajz,Blake:2017ris,Blake:2018leo}. In the current setup, the upper half plane pole skipping point can also be identified by analyzing the equation for $Z_2$ in the sound channel, which is of the same form as \eq{Z eq form}, as will be discussed in section \ref{discussion} and Appendix \ref{stringy upper}.

\section{Gauss-Bonnet correction to pole-skipping}
\label{section GB correction}
\subsection{Setup}

In the above section, we studied the stringy correction which is essentially a fourth order curvature correction $\sim R^4$. In particular, the $\gamma W$ term arises as a top-down correction from a specific string theory (type IIB) to the supergravity action \cite{Grisaru:1986vi,Freeman:1986zh,Park:1987jp,Gross:1986iv}.
This form of correction is just one of a very few known corrections from specific string theories.

Without being restricted to specific known string theory corrections, one may take a pragmatic way to consider generic corrections, usually starting from quadratic curvature corrections
\be
S=\frac{1}{2\kappa_5^2}\int d^5x\sqrt{-g}\left[ R+12+(\alpha_1R^2+\alpha_2R_\mn R^\mn+\alpha_3R_{\mn\rho\sigma}R^{\mn\rho\sigma})\right].
\ee
 {The first two terms of couplings $\alpha_1$ and $\alpha_2$ can be eliminated by a field redefinition of the metric \cite{Kats:2007mq,Brigante:2007nu,Buchel:2008vz}, leaving only the $\alpha_3$ term.
}
The higher curvature terms in general produce higher than second order derivatives in the EOM, and therefore the theory suffers from Ostrogradsky instability and other pathologies \cite{Ostrogradsky:1850fid,Pais:1950za,Stelle:1977ry,Zwiebach:1985uq}. Thus, as the above stringy correction parameterized by $\gamma$, these corrections should only be regarded as perturbations, i.e. $|\alpha_i|\ll 1$. However, for specific combinations of the coefficients, one may obtain the Gauss-Bonnet term (or, the Lovelock term \cite{Lovelock:1971yv} for general higher curvature terms),
\be\label{GB action}
S=\frac{1}{2\kappa_5^2}\int d^5x \sqrt{-g}\left[R+12+\frac{\lambda_{GB}}{2}(R^2-4R_\mn R^\mn+R_{\mn\rho\sigma}R^{\mn\rho\sigma})\right],
\ee
which still leads to second order EOM. Thereby, the theory is expected to circumvent the above difficulties plaguing generic higher curvature theories, and the coupling $\lambda_{GB}$ can be regarded as non-perturbative\footnote{
For example, the KSS bound \cite{Kovtun:2004de} on the shear viscosity to entropy density ratio on CFTs dual to 5D Gauss-Bonnet gravity was obtained for non-perturbative $\lambda_{GB}$ as
$\eta/s=({1}/{4\pi})(1-4\lambda_{GB})$\cite{Brigante:2007nu}.}.

However, the range of $\lambda_{GB}$ is limited to
\be\label{lambdaGB range}
-\frac{7}{36}\leq\lambda_{GB}\leq\frac{9}{100},
\ee
due to causality violation and other issues in the  dual boundary theory \cite{Brigante:2007nu,Brigante:2008gz,Buchel:2009tt}.\footnote{
This constraint on $\lambda_{GB}$ is generalized to general $D$ dimensions with $D\geq 5$ in \cite{Camanho:2009vw,Buchel:2009sk}.
}
Moreover, it was later argued in \cite{Camanho:2014apa} that even for the bulk theory itself, there are bulk causality violation in generic higher curvature gravity, including the Gauss-Bonnet gravity and Lovelock gravity, unless an infinite set of higher spin fields are added.\footnote{
However, see \cite{Papallo:2015rna,Reall:2014pwa} for different opinions.
}
Then the low energy effective theory obtained by integrating out these higher spin fields would modify the action like \eq{GB action} with additional higher derivative terms,  eventually making the EOM higher than second order, and bringing back the difficulties like Ostrogradsky instability. See \cite{Grozdanov:2016fkt} for more detailed discussions. {Besides, there are other instability problems for the Gauss-Bonnet theory, such as the so-called eikonal instability (see \cite{Konoplya:2017zwo} and the references therein). }

Despite the above issues, many features of  the Gauss-Bonnet theory are well-behaved for non-perturbative $\lambda_{GB}$ (at least classically). In particular, exact solutions \cite{Boulware:1985wk,Cai:2001dz} to the second order EOM, and the exact form of the Gibbons-Hawking boundary term \cite{Myers:1987yn} are known. Therefore, we still formally  treat $\lambda_{GB}$ as a non-perturbative parameter in our discussion. Generically, $\lambda_{GB}$ can be regarded as a function of both $\lambda$ and $N_c$. In particular, as a perturbative parameter, it can be interpreted as $\lambda_{GB}\sim 1/N_c$ for $\lambda\gg N_c^{3/2}\gg1$ as in the theory of \cite{Kats:2007mq}. More detailed discussions about the holographic dictionary relating $\lambda_{GB}$ to field theory parameters can be found in \cite{Buchel:2008vz,Buchel:2009sk} as well as \cite{Grozdanov:2018kkt,Grozdanov:2016fkt}.

%As argued in \cite{Kats:2007mq,Grozdanov:2018kkt}, the parameters $\alpha_i$ (and $\lambda_{GB}$ below) can be regarded as proportional to $1/N_c$, although generically it may be a function of both $\lambda$ and $N_c$.

The background solution relevant here is the Gauss-Bonnet black brane \cite{Cai:2001dz}
\be\label{GB black brane}
ds^2=-N_{GB}^2f(r)dt^2+\frac{1}{f(r)}dr^2+r^2(dx^2+dy^2+dz^2),
\ee
where the constant $N_{GB}$ is related to the Gauss-Bonnet coupling $\lambda_{GB}$ as
%\footnote{ {Sometimes it is convenient to define $\gamma_{GB}=\sqrt{1-4\lambda_{GB}}$ instead of $N_{GB}$.}}
\be
N_{GB}^2=\frac{1}{2}(1+\sqrt{1-4\lambda_{GB}}),
\ee
and
\be
f(r)=\frac{r^2}{2\lambda_{GB}}\left(1-\sqrt{1-4\lambda_{GB}(1-\frac{r_0^4}{r^4})}\right).
\ee
 {
In general, for $\lambda_{GB}\leq 1/4$, $N^2_{GB}\geq1/2$.
\footnote{ {Note that at $\lambda_{GB}=1/4$, $N_{GB}^2=1/2$, the shear viscosity vanishes, and the theory exhibits unusual properties in many aspects, such as quasinormal modes and thermodynamics, see \cite{Grozdanov:2016fkt,Chamseddine:1989nu,Crisostomo:2000bb} for detailed discussions. %In particular, the gauge-invariant master fields of the metric perturbations haven been exactly solved in \cite{Grozdanov:2016fkt}. So, using their equations (2.76)-(2.78) to extract the retarded Green's functions from the coefficients of the near boundary expansion, one can easily check that there is no pole-skipping.
Since this value lies far outside of the causality range \eq{lambdaGB range}, we will  not consider it in the following.} }
If, as mentioned above, causality violation  is taken into account, then \eq{lambdaGB range} implies
\be\label{NGB range}
\frac{9}{10}\leq N_{GB}^2\leq\frac{7}{6},
\ee
or approximately, $0.9000\leq N_{GB}^2\leq1.1667$.
}
The temperature of the black brane is
\be\label{GB temperature}
T=N_{GB} \frac{r_0}{\pi}.
\ee
To perform the near horizon analysis, we change to ingoing Eddington-Finkelstein coordinates
\be
v=t+r_*,\ \ \ dr_*=\frac{dr}{N_{GB}f(r)},
\ee
where the metric that we will use takes the form
\be \label{GBBH}
ds^2=-N_{GB}^2f(r)dv^2+2N_{GB}dvdr+r^2(dx^2+dy^2+dz^2).
\ee

\subsection{Scalar field}
Consider a scalar field with mass $m$ determined by the EOM \eq{scalar eom} in the background \eq{GB black brane}. The scalar field EOM becomes
\be\label{scalar eom GB}
 -N_{GB} r^2 f  \phi '' +\phi '  \left(-N_{GB} r^2 f' -3 N_{GB} rf +2 i r^2 \omega \right)+\phi   \left(k^2 N_{GB}+ m^2 N_{GB} r^2+3 i r\omega \right)=0.
\ee
Again, inserting the near horizon expansion of $\phi$ in \eq{scalar NH expansion} into \eq{scalar eom GB} and performing the near horizon expansion lead to \eq{review: order n}, where
the first few coefficients $C_{ij}$ are listed in Appendix \ref{app:GB:scalar} for comparison. We find the similar pattern on the pole-skipping points, i.e., the frequencies $\omega_n=-i2\pi Tn$ exhibit no explicit $N_{GB}$-dependence, while the momenta $k_n$ receive $N_{GB}$-corrections. For example, the first pole-skipping point is
\be
\omega_1=-i2\pi T,\ \ \ k_1^2=-\left(m^2+6\right)r_0^2=-\frac{ \left(m^2+6\right) }{N_{GB}^2}\pi^2T^2.
\ee
The dependence of $k_1^2$ on $r_0$ is the same as  that in the uncorrected case \cite{Blake:2019otz}, whereas the dependence on $N_{GB}$ arises from the relation between $r_0$ and $T$ in \eq{GB temperature}.

Momenta corresponding to $\omega_2$ and $\omega_3$ are given by
\ba
0 &=&k_2^4+2 k_2^2 r_0^2 \left[m^2+4 \left(4 N_{GB}^4-4 N_{GB}^2+3\right)\right]\nonumber\\
 &&+r_0^4 \left[m^4+16 m^2 \left(2 N_{GB}^4-2 N_{GB}^2+1\right)+96 \left(1-2 N_{GB}^2\right)^2\right],\nonumber\\
0&=&8 r_0^2\left[k_3^2+\left(m^2+18\right) r_0^2\right] \left[k_3^2+3 r_0^2 m^2+6r_0^2 \left(-64 N_{GB}^8+128 N_{GB}^6-64 N_{GB}^4+7\right) \right]\nonumber\\
&&-192 \left(m^2+6\right) r_0^6-\left[k_3^2+r_0^2 \left(m^2+96 N_{GB}^4-96 N_{GB}^2+30\right)\right]\left\{\left[k_3^2+\left(m^2+18\right) r_0^2\right]\right.\nonumber\\
 &&\left. \left[k_3^2+r_0^2 \left(m^2+32 N_{GB}^4-32 N_{GB}^2+34\right)\right]-8 r_0^2 \left[k_3^2+3 \left(m^2+12\right) r_0^2\right]\right\},
\ea
 {
from which $k_2^2$ can be solved as
\be
k_2^2= -\left[m^2\pm2 \left(\sqrt{2} \sqrt{m^2+32 N_{GB}^8-64 N_{GB}^6+32 N_{GB}^4+6}+8 N_{GB}^4-8 N_{GB}^2+6\right)\right]r_0^2,
\ee
and $k_3^2$ can also be easily solved, but the expressions are cumbersome and not illuminating, so will not be presented.
}
\subsection{Vector field}

In the Gauss-Bonnet background, the gauge invariant variable $E$ defined in \eq{vector master field} in the shear channel obeys an equation of the same form as \eq{vector eom},
\be\label{GB vector EOM}
E''+A_{E2} E'+B_{E2} E=0,
\ee
with the coefficients given in Appendix \ref{app:GB:vector}.

The leading order near horizon analysis gives the first pole-skipping points
\be
\omega_1=-2\pi Ti, \ \ \ k^2_1=2r_0^2\left(1+r_0\frac{Z'_0}{Z_0}\right)=\frac{2 \pi ^2 T^2}{N_{GB}^3} \left(1+\pi  T \frac{{Z_0'}}{Z_0}\right),
\ee
where $T$ is given by \eq{GB temperature} with higher curvature correction. Similar to the scalar case, we find the same dependence of $k_1^2$ on $r_0$ as that in the uncorrected case \cite{Blake:2019otz}, with the $N_{GB}$-dependence entering through the relation between $r_0$ and $T$ in \eq{GB temperature}.

The momenta corresponding to $\omega_2$ and $\omega_3$ are
\ba
0&=&-k_2^4-8 k_2^2 \left(1-2 N_{GB}^2\right)^2 r_0^2+32 \left(1-2 N_{GB}^2\right)^2 r_0^4+\frac{16 r_0^6 Z_0'^2}{Z_0^2}\nonumber\\
&&+\frac{4r_0^3 \left[k_2^2+4 \left(8 N_{GB}^4-8 N_{GB}^2+1\right) r_0^2\right] Z_0'-16 r_0^6 Z_0''}{Z_0},\\
0&=&k_3^6+2 k_3^4 \left(64 N_{GB}^4-64 N_{GB}^2+21\right) r_0^2-\left[17 k_3^2+18 \left(64 N_{GB}^4-64 N_{GB}^2+7\right) r_0^2\right]4 r_0^6 \frac{Z_0'^2}{Z_0^2}\nonumber\\
&&+576 r_0^9\frac{ Z_0'  Z_0''}{Z_0^2}+12 k_3^2 \left(512 N_{GB}^8-1024 N_{GB}^6+736 N_{GB}^4-224 N_{GB}^2+25\right) r_0^4\nonumber\\
&&+64 r_0^6 \left[k_3^2+9 \left(8 N_{GB}^4-8 N_{GB}^2+1\right) r_0^2\right] \frac{Z_0''}{Z_0}-6 r_0^3 \left[k_3^4+4 k_3^2 \left(32 N_{GB}^4-32 N_{GB}^2+7\right) r_0^2\right.\nonumber\\
&&\left.+12 \left(512 N_{GB}^8-1024 N_{GB}^6+672 N_{GB}^4-160 N_{GB}^2+11\right) r_0^4\right]\frac{ Z_0'}{Z_0}-192 r_0^9 \frac{Z^{'''}_0}{Z_0}\nonumber\\
&&-72 \left(512 N_{GB}^8-1024 N_{GB}^6+736 N_{GB}^4-224 N_{GB}^2+25\right) r_0^6-\frac{360 r_0^9 Z_0'^3}{Z_0^3},\label{k3 GB vector}
\ea
 {
Again, with $Z$ set to unity, in addition to $k_1^2=2r_0^2$, a compact expression can be obtained for the two solutions for $k_2^2$
\be
k_2^2=-4r_0^2 \left(1-2 N_{GB}^2\right)^2\pm 4r_0^2\sqrt{(1-2N_{GB}^2)^2(3-4N_{GB}^2+4N_{GB}^4)}.
\ee
%\ba
%k_2^2&=&-2r_0^2  \left[2\left(1-2 N_{GB}^2\right)^2  -r_0\frac{Z_0'}{Z_0}\right]\pm2r_0^2\sqrt{K}\nonumber\\
%K&\equiv& 4 \left(1-2 N_{GB}^2\right)^2 \left(4 N_{GB}^4-4 N_{GB}^2+3\right) +5r_0^2\frac{Z_0'^2}{Z_0^2} -4\frac{r_0}{Z_0} \left[r_0 Z_0''-4 N_{GB}^2 \left(N_{GB}^2-1\right) Z_0'\right]\nonumber\\
%\ea
It is easy to see that $k_1^2$ is always positive, and the upper ($+$) solution for $k_2^2$ is positive except for $N_{GB}^2=1/2$ where $k_2^2$ vanishes. Moreover, although the expressions for the $k_3^2$ solutions are too lengthy to be listed here, simple numerical analysis of equation \eq{k3 GB vector} (with $Z=1$) indicates that $k_3^2$ always has a positive solution. So there are real solutions for $k_1$, $k_2$ and $k_3$, which correspond to the nontrivial constraints imposed by pole-skipping on the hydrodynamic mode.
}

\subsection{Metric perturbation}
Unlike the theory with stringy correction \eq{Stringy action},  the EOM of Gauss-Bonnet gravity is still of second order. One can simply insert $g_\mn+h_\mn$ into the EOM to obtain the linearized EOM for $h_\mn$ on the black brane background \eq{GBBH}.

As mentioned in Section \ref{sec:stringy:metric perturbation}, The linearized equations decouple according to the symmetry in the plane normal to the direction of propagation, which is taken to be the $x$-direction. Again, we study the perturbations in the shear and sound channels by  inserting corresponding Fourier transform $h_\mn(v,r,x)\rightarrow e^{-i\omega v+ikx}h_\mn(r)$ into the linearized EOM, assuming the radial gauge $h_{r\mu}=0$.

\subsubsection{Shear channel}
In the shear channel, the relevant perturbations are $h_{xy}$ and $h_{vy}$ and the rest are decoupled from them. Following \cite{Kovtun:2005ev}, the gauge invariant master field can be introduced
\be
Z_3=\frac{1}{r^2}(\omega h_{xy}+k h_{vy}),
\ee
which obeys a single second order differential equation
\be\label{Z3 eq}
Z_3''+A_3Z_3'+B_3Z_3=0,
\ee
where $A_3$ and $B_3$ are given in Appendix \ref{app:GB:shear}.

The near horizon analysis leads to pole-skipping points $\omega_{n}=-i2\pi Tn$ and corresponding $k_{n}$.
The momenta corresponding to the first three $\omega_n$ are given from
\ba
0&=&k_1^2 \left(1-2 N_{GB}^2\right)^2+2 \left(8 N_{GB}^4-8 N_{GB}^2-3\right) r_0^2,\nonumber\\
0&=&k_2^4 \left(1-2 N_{GB}^2\right)^2+128 k_2^2 N_{GB}^2 \left(N_{GB}^2-1\right) r_0^2+96 \left(8 N_{GB}^4-8 N_{GB}^2-1\right) r_0^4,\nonumber\\
0&=&k_3^6 \left(1-2 N_{GB}^2\right)^6+6 k_3^4 \left(72 N_{GB}^4-72 N_{GB}^2+5\right) \left(1-2 N_{GB}^2\right)^4 r_0^2\nonumber\\
&&+36 k_3^2 \left(1472 N_{GB}^8-2944 N_{GB}^6+1744 N_{GB}^4-272 N_{GB}^2-5\right) \left(1-2 N_{GB}^2\right)^2 r_0^4\nonumber\\
&&+72 \left(23040 N_{GB}^{12}-69120 N_{GB}^{10}+77760 N_{GB}^8-40320 N_{GB}^6+8952 N_{GB}^4\right.\nonumber\\
&&\left.-312 N_{GB}^2-67\right) r_0^6,\label{k solutions GB shear}
\ea
 {
from which compact expressions can be obtained for $k_1^2$ and $k_2^2$ as
\ba
k_1^2&=&\frac{2r_0^2 \left(3+8 N_{GB}^2-8 N_{GB}^4\right)}{\left(1-2 N_{GB}^2\right)^2}, \label{GB shear k1 sol}\\
k_2^2&=&\frac{4r_0^2 \left(16 N_{GB}^2-16 N_{GB}^4\pm\sqrt{2} \sqrt{32 N_{GB}^8-64 N_{GB}^6+20 N_{GB}^4+12 N_{GB}^2+3}\right)}{\left(1-2 N_{GB}^2\right)^2}.\label{k2 GB}
\ea
}

 {
To ensure the existence of a real solution for $k_1$, one must have $3+8 N_{GB}^2-8 N_{GB}^4>0$, which implies $N_{GB}^2<(2+\sqrt{10})/4\approx1.2906$. Compared with the range arising from the argument of causality violation, one can see that \eq{NGB range} ensures that $k_1$ has a real solution, corresponding to the hydrodynamic diffusion mode. An easy analytic analysis of \eq{k2 GB} indicates that the upper ($+$) branch of the $k_2^2$ solutions can give real $k_2$ for $N_{GB}^2<(2+\sqrt{6})/4\approx1.1124$. So this can be regarded as an upper bound which is tighter than the causality upper bound in \eq{NGB range}, if the existence of real solutions for $k_2$ is required {\it a priori}. Furthermore, numerical analysis of the equation for $k_3$ in \eq{k solutions GB shear} suggests that $k_3$ can be real for $N_{GB}^2<1.0632$, which is an even tighter upper bound.
Based on these observations, one can expect that, in general, requiring the existence of real solutions for $k_n$ imposes a $n$-dependent upper bound on $N_{GB}^2$, and that this bound becomes tighter for larger $n$. Moreover, $k_n$ approaches zero for $N_{GB}^2$ approaching its upper bound for $n$, such that at this particular pole-skipping point, $|\omega_n|\sim T$, but $|k_n|\ll T$. This is in contrast to the generic pole-skipping  phenomenon without higher curvature corrections, where {\it both} $|\omega_n|\sim T$ and $|k_n|\sim T$ \cite{Grozdanov:2017ajz,Blake:2018leo,Blake:2019otz}. In addition, it would be interesting to further explore the physical implication when $k_n$ has no real solutions but $N_{GB}$ is still within the range in \eq{NGB range}.
}

\subsubsection{Sound channel}

In the sound channel, the gauge invariant variable constructed using the relevant perturbations is given by
\be
Z_4=\frac{1}{r^2}\left[ k^2 h_{vv}+\omega^2h_{xx}+2\omega k h_{vx}+\left(\frac{N_{GB}^2f'}{2r}k^2-\omega^2\right)\frac{h_{aa}}{2} \right].
\ee
The equation for $Z_4$ is again of the form
\be\label{Z4 eq}
Z_4''+A_4Z_4'+B_4Z_4=0,
\ee
where $A_4$ and $B_4$ are given in Appendix \ref{app:GB:sound}.

The near horizon analysis in this case again gives the pole-skipping frequencies $\omega_n=-i2\pi n T$. The momenta corresponding to the first three $\omega_n$ are
\ba
0&=&k_1^4 \left(-8 N_{GB}^4+8 N_{GB}^2-1\right) r_0+4 k_1^2 \left(8 N_{GB}^4-8 N_{GB}^2+1\right) r_0^3+12 r_0^5 \left(8 N_{GB}^4-8 N_{GB}^2-3\right),\nonumber\\
0&=&k_2^4 \left(8 N_{GB}^4-8 N_{GB}^2+1\right)^2 r_0^2-8 k_2^2 \left(64 N_{GB}^8-128 N_{GB}^6+72 N_{GB}^4-8 N_{GB}^2+1\right) r_0^4\nonumber\\
&&-96 \left(1-2 N_{GB}^2\right)^2 \left(8 N_{GB}^4-8 N_{GB}^2-1\right) r_0^6,\nonumber\\
0&=&-k_3^6 \left(8 N_{GB}^4-8 N_{GB}^2+1\right)^3 r_0^3-2 k_3^4 \left(4608 N_{GB}^{12}-13824 N_{GB}^{10}+16576 N_{GB}^8-10112 N_{GB}^6\right.\nonumber\\
&&\left.+3096 N_{GB}^4-344 N_{GB}^2+9\right) r_0^5+4 k_3^2 \left(59904 N_{GB}^{12}-179712 N_{GB}^{10}+199104 N_{GB}^8\right.\nonumber\\
&&\left.-98688 N_{GB}^6+20536 N_{GB}^4-1144 N_{GB}^2+37\right) r_0^7+72 \left(23040 N_{GB}^{12}-69120 N_{GB}^{10}\right.\nonumber\\
&&\left.+77760 N_{GB}^8-40320 N_{GB}^6+8952 N_{GB}^4-312 N_{GB}^2-67\right) r_0^9,\label{GB sound k}
\ea
 {
from which compact expressions can be obtained for $k^2_1$ and $k^2_2$  as
\ba
k_1^2&=&2 r_0^2\pm\frac{4 \sqrt{2} r_0^2\sqrt{32 N_{GB}^8-64 N_{GB}^6+28 N_{GB}^4+4 N_{GB}^2-1 }}{8 N_{GB}^4-8 N_{GB}^2+1},\\
k_2^2&=&\frac{4  r_0^2\left(64 N_{GB}^8-128 N_{GB}^6+72 N_{GB}^4-8 N_{GB}^2+1\right)\pm4 r_0^2\sqrt{K }}{\left(8 N_{GB}^4-8 N_{GB}^2+1\right)^2},
\ea
with
\ba
K&\equiv&\left(16384 N_{GB}^{16}-65536 N_{GB}^{14}+103936 N_{GB}^{12}-82432 N_{GB}^{10}+33664 N_{GB}^8-6400 N_{GB}^6\right.\nonumber\\
&&\left.+328 N_{GB}^4+56 N_{GB}^2-5\right).
\ea
$k_1^2$ diverges for $N^2_{GB}=(2+\sqrt{2})/4\approx0.8536$,\footnote{
This is also noted in \cite{overlapping} in their Appendix C, at the corresponding $\lambda_{GB}=1/8$.
} implying that no pole-skipping occurs at this point. Note that this value of $N_{GB}^2$ is outside the range given in \eq{NGB range}. In contrast, it is not hard to see, by inserting the above $N_{GB}^2$ value into \eq{GB sound k}, that $k_2^2$ and $k_3^2$ always have finite solutions.}

Again, the upper half plane pole-skipping location can be extracted from equation \eq{Z4 eq}, as will be discussed in the next section and Appendix \ref{GB upper}.
\section{Discussion}
\label{discussion}
In this paper, we have studied the effect of the stringy correction ($\sim R^4$) and the Gauss-Bonnet correction $(\sim R^2)$ to the pole-skipping phenomenon of typical scalar, vector and tensor operators dual to corresponding bulk fields. Of course, as one can easily check, all of our results recover the known results in the uncorrected case previously studied in \cite{Blake:2019otz}. Something new here is that, the general feature of these pole-skipping points, i.e., the locations of the frequencies are all given by $\omega_{n}=-i2\pi Tn$, with the corrections only modify the expression of the temperature. On the other hand, the momenta $k_{n}$ receive explicit stringy or Gauss-Bonnet corrections. The similarity in this qualitative feature for these higher curvature corrections  is in keeping with the results discussed in \cite{Grozdanov:2016vgg}, where the quasinormal spectra of metric perturbations are shown to exhibit similar behavior regardless of the $R^2$ and $R^4$ corrections.

Moreover, the way these corrections affect the frequencies and momenta is similar to the pole-skipping point of chaos in the upper half complex $\omega$ plane as studied in \cite{Grozdanov:2018kkt}. For example, there, pole-skipping occurs at $\omega_*=+i2\pi T$, and $k_*=i\sqrt{6}\pi T(1-\gamma 23/2)$ for the stringy correction, and $k_*=i\sqrt{6}\pi T/N_{GB}$ for the Gauss-Bonnet correction. It is in this way that the butterfly velocity $v_B=\omega_*/k_*$ receives correction. This suggests that at the pole-skipping points, the dependence of frequency on temperature exhibits certain universality,  {which is robust against the finite $N_c$ and finite 't Hooft coupling corrections which are holographically dual to the typical $R^2$ and $R^4$ corrections studied here and in \cite{Grozdanov:2018kkt}. Of course, it would be interesting to further investigate the robustness of this universality under more general higher order curvature corrections.
}

In fact, the upper half plane pole-skipping point can also be obtained by studying a special point of the sound channel equations \eq{Z eq form} and \eq{Z4 eq}, following the argument given in \cite{Blake:2019otz} for the uncorrected case. The special point here refers to a particular relation between $k$ and $\omega$ at which the pole structures of the coefficients $A$ and $B$ in \eq{Z eq form} or \eq{Z4 eq} change. The technical details of calculation are given in Appendix \ref{stringy upper} and \ref{GB upper}. Our results   indicates that, instead of analyzing the $vv$ component of near horizon Einstein equation, the pole-skipping point of chaos in the upper   half plane of complex $\omega$ can also be obtained by analyzing the equation for the gauge invariant variables in the sound channel, even in the presence of typical $R^2$ and $R^4$ higher curvature corrections. This further lends support to the expectation that pole-skipping is a universal phenomenon holographically encoded by near horizon physics.

 {
In the absence of any higher curvature correction, it was argued in \cite{Grozdanov:2019uhi} that two important parameters of chaos, i.e. $\lambda_L$ and $v_B$, can be recovered, {\it irrespectively of the channel} of metric perturbations, as
%\be
%\lambda_L=2\pi T|w_*|,\ \ \ v_B=\frac{|w_*|}{|q_*|}
%\ee
%where the dimensionless frequencies $w\equiv\omega/2\pi T$ and momenta $q\equiv k/2\pi T$ in the sound, shear and scalar channels are, respectively,
%\ba
%w_*&=&i,\ -i,\ -i,\\
%q_*&=&\sqrt{\frac{3}{2}}i,\ \sqrt{\frac{3}{2}},\ \sqrt{\frac{3}{2}}i,
%\ea
\be\label{universal expression}
\lambda_L=|\omega_*|,\ \ \ v_B=\frac{|\omega_*|}{|k_*|}
\ee
where $\omega_*$ and $k_*$ in different channels are,
\ba
    &{\rm sound\ channel}&: \omega_*=+i2\pi T,\ \ \  k_*=i\sqrt{6}r_0,\\
    &{\rm shear\ channel}&: \omega_* =-i2\pi T,\ \ \  k_*=\sqrt{6}r_0,\\
    &{\rm scalar\ channel}&:  \omega_*=-i2\pi T,\ \ \  k_*=i\sqrt{6}r_0.
\ea
Note that the results in the last two channels are just the first pole-skipping points $(\omega_1,k_1)$.\footnote{
In the shear channel, this can be checked by setting $\gamma=0$ in \eq{stringy shear k1 sol} or $N_{GB}=1$ in \eq{GB shear k1 sol}. Although the results in the scalar channel are not presented in this paper, the calculations have been performed, and the results, in particular $(\omega_1,k_1)$, have been checked.}
In the presence of the higher curvature corrections studied here, $\omega_*$ remains the same, but $k_*$ changes differently in the three channels. From the detailed results listed in Appendix \ref{app:corrections to k1}, one finds that only the results (c.f. Appendix \ref{stringy GB upper}) of pole-skipping in the {\it upper} half plane in the sound channel agree with the shockwave analysis of the OTOC \cite{Grozdanov:2018kkt}. So this suggests that, when the higher curvature corrections are taken into account, $\lambda_L$ in the united expression in \eq{universal expression} still holds for all channels, whereas $v_B$ can only be obtained from the sound channel, not from the other two channels.
}

 {
We end this paper by noting an interesting open question worthy of further investigation. In the study of the gravitational quasinormal modes in the presence of higher curvature corrections, it has been found that there is a series of modes with pure imaginary frequencies which are non-perturbative in $\gamma$ or $\lambda_{GB}$, and absent in the Einstein gravity limit at $\gamma=0$ or $\lambda_{GB}=0$ \cite{Grozdanov:2016vgg} (see also \cite{Grozdanov:2016fkt,Konoplya:2017zwo,Konoplya:2017ymp,Gonzalez:2017gwa} for related discussions). In particular, in the shear channel, in addition to the gapless hydrodynamic diffusion mode, there exist other modes having pure imaginary frequencies in the lower half plane with real momenta. Moreover, the first of these non-perturbative modes can interact with the hydrodynamic  mode when they are colliding at certain critical value $\gamma_c$ or $\lambda_{GB}^c$ for a fixed momentum (or, equivalently, at certain momentum $k_c$ for fixed $\gamma$ or $\lambda_{GB}$), after which the latter acquires real parts and the hydrodynamic description breaks down. Since the pole-skipping in the shear channel studied in this paper also occurs in the lower half plane, and it has been found that pole-skipping imposes nontrivial constraints on the hydrodynamic mode, it would be interesting to study the relation between the non-perturbative modes and the pole-skipping points. %In contrast to the method use here, where $\omega$ is fixed to $-in2\pi T$ and then real $k_n$ is solved, one can search for pure imaginary $\omega$ for some fixed real $k$, as was usually done in the literature such as \cite{Grozdanov:2016vgg}.
This requires high precision numerical methods. Hopefully, the results will be reported in a future publication.
}

After the completion of this paper, \cite{overlapping} appears in arXiv, which has some overlapping with the discussion here  of the Gauss-Bonnet correction. It can be checked that the results agree where they overlap.

\acknowledgments

The calculations performed in this work were facilitated by two Mathematica packages: {\it xAct}, in particular, {\it xCoba} by David Yllanes and Jos\'{e} M. Mart\'{i}n-Garc\'{i}a,
%available at {www.xact.es}, 
and {\it RGTC} by Sotirios Bonanos.
%available at {www-old.inp.demokritos.gr/~sbonano//RGTC/}. 
This work was in part supported by the NSFC grant 11647088, and a 333 grant from the North University of China.

\appendix

\section{Stringy correction to scalar field}
%\subsection{Source term in scalar EOM}
\label{app:stringy:scalar}
The $\gamma$-dependent source term in the scalar EOM \eq{scalar eom fourier} is
\be
S_1=\frac{15 r_0^4}{r^{12}}(S_{10}\phi+S_{11}\phi'+S_{12}\phi''),
\ee
where
\ba
S_{10}&=&-8 r r_0^8 \left(k^2+m^2 r^2\right), \nonumber\nonumber\\
S_{11}&=&112 r^4 r_0^8+5 r^{12}-121 r_0^{12}, \nonumber\nonumber\\
S_{12}&=&r (11 r_0^{12} - 16 r_0^8 r^4 + 5 r^{12}).
\ea
The coefficients of the equations \eq{review: order n} with stringy corrections are
\ba
C_{10}&=&-k^2-3 r_0 \left(m^2 r_0+2 i \omega \right)-120 \gamma  \left(11 k^2+9 m^2 r_0^2\right),\nonumber\\
C_{11}&=&-r_0 \left[k^2+  m^2r_0^2-20  r_0^2+9 i \omega r_0\right] +60 \gamma  \left[2 k^2 r_0+\left(2 m^2-139\right) r_0^3\right],\nonumber\\
C_{12}&=&=4 r_0^3 (4 r_0-i \omega )+240 \gamma  r_0^4,\nonumber\\
C_{20}&=&-3 \left(m^2 r_0+i \omega \right)+\gamma  \left(\frac{7920 k^2}{r_0}+5400 m^2 r_0\right),\nonumber\\
C_{21}&=&-k^2-3 r_0 \left[\left(m^2-10\right) r_0+4 i \omega \right]-30 \gamma  \left[44 k^2+3 \left(12 m^2-901\right) r_0^2\right],\nonumber\\
C_{22}&=&-r_0 \left[k^2+ \left(m^2-60\right) r_0^2+15 i \omega r_0\right] +60 \gamma  \left[2 k^2 r_0+\left(2 m^2-417\right) r_0^3\right],\nonumber\\
C_{23}&=&6 r_0^3 (6 r_0-i \omega )+540 \gamma  r_0^4.
\ea

\section{Stringy corrections to vector field}
\label{app:stringy:vector}
The  coefficients $A_{E1}$ and $B_{E1}$ in equation \eq{stringy vector EOM} are given from
\ba
\alpha_E A_{E}&=&\left\{-r \omega ^2 Z  Z_{vv}  f' +k^2 f ^2 Z_{vv} ^2 \left(r Z' +3 Z \right)-\omega ^2 f  \left[r Z_{vv}  Z' +Z  \left(r Z_{vv}' +3 Z_{vv} \right)\right]\right\}\nonumber\\
&&-\frac{2 i \omega  Z_{vr} }{r^2 f  Z_{vv} }-\frac{Z_{vr}' }{Z_{vr} },\\
\beta_E B_{E}&=&i Z_{vr}  \left\{r^2 \omega  Z'  \left(\omega ^2-k^2 f  Z_{vv} \right)+Z  \left[r \omega  \left(k^2 r Z_{vv}  f' +k^2 f  (r Z_{vv}' -Z_{vv}  )+\omega ^2\right)\right.\right.\nonumber\\
&&\left.\left.+i Z_{vr}  \left(k^4 f  Z_{vv} -k^2 \omega ^2\right)\right]\right\},
\ea
where
\ba
\alpha_E&=&rf  Z  Z_{vv}  \left(k^2 f  Z_{vv} -\omega ^2\right),\nonumber\\
\beta_E&=&r^3\alpha_E.
\ea

\section{Stringy corrections to metric perturbations}

\subsection{Derivation of the equation for the master field of metric perturbations}
\label{app:derivation:master equation}
In coordinates other than ingoing Eddington-Finkelstein coordinates, the equations for the master fields in the shear and sound channels have been discussed in \cite{Buchel:2004di,Benincasa:2005qc,Buchel:2008bz,Stricker:2013lma,Grozdanov:2016vgg,Solana:2018pbk,Buchel:2018eax}. The basic form of the equations is
\be
Z''+AZ'+BZ=\gamma (M_0Z+M_1Z').
\ee
Or, equivalently, inserting $Z={Z^{(0)}}+\gamma {Z^{(1)}}$ leads to
\ba
\co(\gamma^0):&&{Z^{(0)}}''+A{Z^{(0)}}'+B{Z^{(0)}}=0, \\
\co(\gamma):&&{Z^{(1)}}''+A{Z^{(1)}}'+B{Z^{(1)}}=M_0{Z^{(0)}}+M_1{Z^{(0)}}'.
\ea
Here we present the basic strategy to derive the equations in ingoing Eddington-Finkelstein coordinates for our discussion of pole-skipping.

The EOM of the perturbations $h_\mn$ are of the form
\be
\Psi''+a\Psi'+b\Psi=\gamma G[\Psi^{''''},\Psi^{'''},\Psi'',\Psi',\Psi],
\ee
where  $\Psi$ denotes $h_\mn$ for notational simplicity, and the $\gamma$-dependent source $G$ involves higher derivatives arising from the stringy correction $\gamma W$ in \eq{Stringy action}. Inserting $\Psi={\Psi^{(0)}}+\gamma {\Psi^{(1)}}$, the above equation can also be written as
\ba
\co(\gamma^0):&&{\Psi^{(0)}}''+a{\Psi^{(0)}}'+b{\Psi^{(0)}}=0,\label{master target eq 0}\\
\co(\gamma):&&{\Psi^{(1)}}''+a{\Psi^{(1)}}'+b{\Psi^{(1)}}= G[\Psi^{(0)''''},{\Psi^{(0)'''}},\Psi''_0,{\Psi^{(0)}}',{\Psi^{(0)}}].\label{master target eq 1}
\ea

The equation for the master field can be obtained as follows
\bnum
    \item
   As discussed in \cite{Solana:2018pbk,Buchel:2018eax}, we can use \eq{master target eq 0} to substitute the higher derivatives in $G$ in terms of $\Psi'_0$ and ${\Psi^{(0)}}$. Then \eq{master target eq 1} becomes
    \be \label{sound target eq 0 new}
    {\Psi^{(1)}}''+a{\Psi^{(1)}}'+b{\Psi^{(1)}}=m_0 {\Psi^{(0)}}+m_1 {\Psi^{(0)}}'
    \ee
    \item
    Insert the expression for the master field\footnote{For example, in the shear channel \eq{stringy:shear:Z1}, $\Psi_i$ are $h_{xy}$ and $h_{vy}$, with corresponding coefficients $\alpha_i$ as $\omega/r^2$ and $k/r^2$, respectively.
    }
    \be
    Z=\sum\alpha_i\Psi_i
    \ee
     into the ansatz
    \be \label{ab ansatz}
     {Z^{(1)}}''+A{Z^{(1)}}'+B{Z^{(1)}}= M_0 {Z^{(0)}}+M_1 {Z^{(0)}}',
    \ee
    where $A,B,M_0$ and $M_1$ are to be determined.
    \item
    Using \eq{sound target eq 0 new} to replace all $\Psi_{\alpha1}''$, such that \eq{ab ansatz} takes the form
    \[
    \alpha_1\Psi_{i1}+\alpha_2\Psi_{i1}'+\alpha_3 \Psi_{i0}+\alpha_4 \Psi_{i0}'=0,
    \]
    where the four coefficients $\alpha_i$ are functions of $A,B,M_0$ and $M_1$. The vanishing of all $\alpha_i$'s gives four algebraic equations to solve for $A,B,M_0$ and $M_1$.
\enum

%Moreover, one can also combine $h_\mn$ to obtain the equation for the master field $Z$, which also take the above form. And then perform the analysis \eq{} to \eq{} to obtain the desired equation \eq{}.

\subsection{Stringy corrections  in the shear channel}
\label{app:stringy:shear}
In the shear channel, the coefficients in \eq{Z eq form} are
\ba
A&=&\frac{k^2 \left(r^4-r_0^4\right) \left(5 r^4-2 i r^3 \omega -5 r_0^4\right)+r^4 \omega ^2 \left(-5 r^4+2 i r^3 \omega +r_0^4\right)}{r \left(r^4-r_0^4\right) \left[k^2 \left(r^4-r_0^4\right)-r^4 \omega ^2\right]},\\
B&=&\frac{k^4 \left(r_0^4-r^4\right)+k^2 r \omega  \left(-3 i r^4+r^3 \omega +7 i r_0^4\right)+3 i r^5 \omega ^3}{\left(r^4-r_0^4\right) \left[k^2 \left(r^4-r_0^4\right)-r^4 \omega ^2\right]},
\ea
\ba
M_0&=&\frac{-r_0^4}{r^{12} \left(r^4-r_0^4\right) \left(k^2 \left(r^4-r_0^4\right)-r^4 \omega ^2\right)^2}
\left[48 k^8 r_0^4 \left(r^5-r r_0^4\right)^2\right.\nonumber\\
&&\left.-k^6 \left(r^4-r_0^4\right) \left(75 r^{12}-1440 r^8 r_0^4-640 i r^7 r_0^4 \omega +96 r^6 r_0^4 \omega ^2+2640 r^4 r_0^8+640 i r^3 r_0^8 \omega\right.\right.\nonumber\\
&&\left.\left.-1275 r_0^{12}\right)+k^4 r \omega  \left(75 i r^{16}+150 r^{15} \omega +3405 i r^{12} r_0^4-3744 r^{11} r_0^4 \omega -800 i r^{10} r_0^4 \omega ^2\right.\right.\nonumber\\
&&\left.\left.+48 r^9 r_0^4 \omega ^3-10080 i r^8 r_0^8+7296 r^7 r_0^8 \omega +992 i r^6 r_0^8 \omega ^2+9585 i r^4 r_0^{12}-3702 r^3 r_0^{12} \omega\right.\right.\nonumber\\
&&\left.\left.-2985 i r_0^{16}\right)+i k^2 r^5 \omega ^3 \left(150 r^{12}+75 i r^{11} \omega -4908 r^8 r_0^4-3093 i r^7 r_0^4 \omega +160 r^6 r_0^4 \omega ^2\right.\right.\nonumber\\
&&\left.\left.+8916 r^4 r_0^8+3291 i r^3 r_0^8 \omega -4158 r_0^{12}\right)-9 i r^9 \omega ^5 \left(25 r^8-167 r^4 r_0^4-96 i r^3 r_0^4 \omega -7 r_0^8\right)\right]\nonumber\\
\ea
\ba
M_1&=&\frac{2r_0^4}{r^{13} \left(r^4-r_0^4\right) \left[k^2 \left(r^4-r_0^4\right)-r^4 \omega ^2\right]^2}
\left[320 k^6 r^2 r_0^4 \left(r^4-r_0^4\right)^3\right.\nonumber\\
&&\left.+k^4 \left(r^4-r_0^4\right) \left(75 i r^{15} \omega +1440 r^{12} r_0^4-400 r^{10} r_0^4 \omega ^2-3960 r^8 r_0^8-240 i r^7 r_0^8 \omega \right.\right.\nonumber\\
&&\left.\left.+496 r^6 r_0^8 \omega ^2+3600 r^4 r_0^{12}+165 i r^3 r_0^{12} \omega -1080 r_0^{16}\right)+2 k^2 r^4 \omega ^2 \left(r^4-r_0^4\right) \left(75 r^{12}\right.\right.\nonumber\\
&&\left.\left.-75 i r^{11} \omega -1002 r^8 r_0^4-75 i r^7 r_0^4 \omega +40 r^6 r_0^4 \omega ^2+1374 r^4 r_0^8+165 i r^3 r_0^8 \omega -462 r_0^{12}\right)\right.\nonumber\\
&&\left.+3 r^8 \omega ^4 \left(-50 r^{12}+25 i r^{11} \omega +238 r^8 r_0^4+25 i r^7 r_0^4 \omega +70 r^4 r_0^8-55 i r^3 r_0^8 \omega -258 r_0^{12}\right)\right].\nonumber\\
\ea

\subsection{Stringy corrections   in the sound channel}
\label{app:stringy:sound}
In the sound channel, the coefficients in \eq{Z eq form} are
\ba
A&=&\frac{k^2 \left(15 r^8-6 i r^7 \omega -16 r^4 r_0^4+2 i r^3 r_0^4 \omega +9 r_0^8\right)+3 r^4 \omega ^2 \left(-5 r^4+2 i r^3 \omega +r_0^4\right)}{r \left(r^4-r_0^4\right) \left[k^2 \left(3 r^4-r_0^4\right)-3 r^4 \omega ^2\right]},\\
B&=&\frac{k^4 \left(r^2 r_0^4-3 r^6\right)+k^2 \left(3 r^6 \omega  (\omega -3 i r)+11 i r^3 r_0^4 \omega +16 r_0^8\right)+9 i r^7 \omega ^3}{r^2 \left(r^4-r_0^4\right) \left[k^2 \left(3 r^4-r_0^4\right)-3 r^4 \omega ^2\right]},
\ea
\ba
M_{0}&=&-\frac{r_0^4}{r^{14} \left(r^4-r_0^4\right) \left[k^2 \left(3 r^4-r_0^4\right)-3 r^4 \omega ^2\right]^3} \left\{48 k^{10} r^4 r_0^4 \left(3 r^4-r_0^4\right)^3\right.\nonumber\\
&&\left.-k^8 r^2 \left(3 r^4-r_0^4\right) \left(675 r^{16}-35199 r^{12} r_0^4-6240 i r^{11} r_0^4 \omega +1296 r^{10} r_0^4 \omega ^2+74004 r^8 r_0^8\right.\right.\nonumber\\
&&\left.\left.+5952 i r^7 r_0^8 \omega -432 r^6 r_0^8 \omega ^2-41287 r^4 r_0^{12}-1120 i r^3 r_0^{12} \omega +5811 r_0^{16}\right)\right.\nonumber\\
&&\left.+3 k^6 \left[-225 i r^{23} \omega +2025 r^{22} \omega ^2+7200 r^{20} r_0^4+72048 i r^{19} r_0^4 \omega -92637 r^{18} r_0^4 \omega ^2\right.\right.\nonumber\\
&&\left.\left.-12960 i r^{17} r_0^4 \omega ^3+432 r^{16} \left(544 r_0^8+3 r_0^4 \omega ^4\right)-216363 i r^{15} r_0^8 \omega +188316 r^{14} r_0^8 \omega ^2\right.\right.\nonumber\\
&&\left.\left.+12992 i r^{13} r_0^8 \omega ^3-16 r^{12} r_0^8 \left(51884 r_0^4+27 \omega ^4\right)+252721 i r^{11} r_0^{12} \omega -107253 r^{10} r_0^{12} \omega ^2\right.\right.\nonumber\\
&&\left.\left.-2720 i r^9 r_0^{12} \omega ^3+895808 r^8 r_0^{16}-113384 i r^7 r_0^{16} \omega +17209 r^6 r_0^{16} \omega ^2-360288 r^4 r_0^{20}\right.\right.\nonumber\\
&&\left.\left.+15019 i r^3 r_0^{20} \omega +44736 r_0^{24}\right]+3 k^4 r^4 \omega ^2 \left(2475 i r^{19} \omega -2025 r^{18} \omega ^2-14400 r^{16} r_0^4\right.\right.\nonumber\\
&&\left.\left.-116751 i r^{15} r_0^4 \omega +86778 r^{14} r_0^4 \omega ^2+7200 i r^{13} r_0^4 \omega ^3-144 r^{12} \left(2713 r_0^8+3 r_0^4 \omega ^4\right)\right.\right.\nonumber\\
&&\left.\left.+259788 i r^{11} r_0^8 \omega -126558 r^{10} r_0^8 \omega ^2-4960 i r^9 r_0^8 \omega ^3+1276224 r^8 r_0^{12}-258063 i r^7 r_0^{12} \omega\right.\right.\nonumber\\
&&\left.\left.+41323 r^6 r_0^{12} \omega ^2-1118016 r^4 r_0^{16}+79419 i r^3 r_0^{16} \omega +273264 r_0^{20}\right)\right.\nonumber\\
&&\left.+9 k^2 r^8 \omega ^4 \left(-1425 i r^{15} \omega +225 r^{14} \omega ^2+2400 r^{12} r_0^4+19410 i r^{11} r_0^4 \omega -12447 r^{10} r_0^4 \omega ^2\right.\right.\nonumber\\
&&\left.-160 i r^9 r_0^4 \omega ^3+51888 r^8 r_0^8-14286 i r^7 r_0^8 \omega +7993 r^6 r_0^8 \omega ^2-148560 r^4 r_0^{12}+8187 i r^3 r_0^{12} \omega \right.\nonumber\\
&&\left.\left.+86832 r_0^{16}\right)+243 i r^{15} \omega ^7 \left(25 r^8-167 r^4 r_0^4-96 i r^3 r_0^4 \omega -7 r_0^8\right)\right\},
\ea
\ba
M_1&=& \frac{2 r_0^4 }{r^{13} \left(r^4-r_0^4\right) \left[k^2 \left(3 r^4-r_0^4\right)-3 r^4 \omega ^2\right]^3}\left[16 k^8 r^2 r_0^4 \left(585 r^{16}-1338 r^{12} r_0^4+1044 r^8 r_0^8\right.\right.\nonumber\\
&&\left.\left.-326 r^4 r_0^{12}+35 r_0^{16}\right)-3 k^6 \left(450 r^{24}-675 i r^{23} \omega -34224 r^{20} r_0^4+6480 r^{18} r_0^4 \omega ^2\right.\right.\nonumber\\
&&\left.\left.+133998 r^{16} r_0^8+1935 i r^{15} r_0^8 \omega -12976 r^{14} r_0^8 \omega ^2-221832 r^{12} r_0^{12}-1685 i r^{11} r_0^{12} \omega \right.\right.\nonumber\\
&&\left.\left.+7856 r^{10} r_0^{12} \omega ^2+177430 r^8 r_0^{16}+520 i r^7 r_0^{16} \omega -1360 r^6 r_0^{16} \omega ^2-63304 r^4 r_0^{20}-55 i r^3 r_0^{20} \omega\right.\right.\nonumber\\
&&\left.\left.+7482 r_0^{24}\right)+3 k^4 r^4 \omega ^2 \left(2250 r^{20}-2025 i r^{19} \omega -54438 r^{16} r_0^4-675 i r^{15} r_0^4 \omega +3600 r^{14} r_0^4 \omega ^2\right.\right.\nonumber\\
&&\left.\left.+159972 r^{12} r_0^8+5580 i r^{11} r_0^8 \omega -6080 r^{10} r_0^8 \omega ^2-227448 r^8 r_0^{12}-3195 i r^7 r_0^{12} \omega \right.\right.\nonumber\\
&&\left.\left.+2480 r^6 r_0^{12} \omega ^2+158466 r^4 r_0^{16}+495 i r^3 r_0^{16} \omega -38802 r_0^{20}\right)-9 k^2 r^8 \omega ^4 \left(1050 r^{16}\right.\right.\nonumber\\
&&\left.\left.-675 i r^{15} \omega -8880 r^{12} r_0^4-450 i r^{11} r_0^4 \omega +80 r^{10} r_0^4 \omega ^2+8028 r^8 r_0^8+1710 i r^7 r_0^8 \omega\right.\right.\nonumber\\
&&\left.\left.-80 r^6 r_0^8 \omega ^2-2724 r^4 r_0^{12}-495 i r^3 r_0^{12} \omega +2526 r_0^{16}\right)+81 r^{12} \omega ^6 \left(50 r^{12}-25 i r^{11} \omega\right.\right.\nonumber\\
&&\left.\left. -238 r^8 r_0^4-25 i r^7 r_0^4 \omega -70 r^4 r_0^8+55 i r^3 r_0^8 \omega +258 r_0^{12}\right) \right].
\ea

\section{Gauss-Bonnet corrections to scalar field}

\label{app:GB:scalar}

The coefficients of the equations \eq{review: order n} with Gauss-Bonnet corrections are
\ba
C_{10}&=&-k^2 N_{GB}-3 r_0 \left(m^2 N_{GB} r_0+2 i \omega \right),\nonumber\\
C_{11}&=&-r_0 \left[k^2 N_{GB}+\left(m^2-20\right) N_{GB} r_0^2+32 N_{GB}^5 r_0^2-32 N_{GB}^3 r_0^2+9 i \omega r_0  \right],\nonumber\\
C_{12}&=&4 r_0^3 (4 N_{GB} r_0-i \omega ),\nonumber\\
C_{20}&=&-3 \left(m^2 N_{GB} r_0+i \omega \right),\nonumber\\
C_{21}&=&-k^2 N_{GB}-3 r_0 \left[\left(m^2-10\right) N_{GB} r_0-128 N_{GB}^9 r_0+256 N_{GB}^7 r_0-128 N_{GB}^5 r_0+4 i \omega \right],\nonumber\\
C_{22}&=&-r_0 \left[k^2 N_{GB}+\left(m^2-60\right) N_{GB} r_0^2+96 N_{GB}^5 r_0^2-96 N_{GB}^3 r_0^2+15 i \omega r_0\right],\nonumber\\
C_{23}&=&6 r_0^3 (6 N_{GB} r_0-i \omega ).
\ea
\section{Gauss-Bonnet corrections to vector field}
\label{app:GB:vector}
The coefficients of the equation \eq{GB vector EOM} for the gauge invariant variable are
\ba
A_{E2}&=&\frac{Z' }{Z }+\frac{r^3 \omega ^2 \left(-N_{GB} f' +2 i \omega \right)+3 k^2 N_{GB}^3 f ^2+N_{GB} r \omega  f  \left(-r \omega -2 i k^2 N_{GB}\right)}{N_{GB} r f  \left(k^2 N_{GB}^2 f -r^2 \omega ^2\right)},\nonumber\\
B_{E2}&=&\frac{1}{N_{GB} r^2 f   \left(k^2 N_{GB}^2 f -r^2 \omega ^2\right)}\left\{  r^2 \omega  \left[i k^2 N_{GB}^2 f' +k^2 N_{GB} \omega +i r \omega ^2\right.\right. \nonumber\\
&&\left.\left.-k^2 N_{GB}^2 f  \left(k^2 N_{GB}+3 i r \omega \right)\right]+i r^2 \omega  \frac{Z'}{Z}  \left(r^2 \omega ^2-k^2 N_{GB}^2 f \right)\right\}.
\ea
\section{Gauss-Bonnet corrections to metric perturbations}
\subsection{Gauss-Bonnet corrections   in the shear channel}
\label{app:GB:shear}
The coefficients in the equation for $Z_3$ \eq{Z3 eq} are
\ba
i\alpha_3 A_3&=&
i N_{GB} r^7 \omega  f  \left[2 i k^2 N_{GB} \left(1-2 N_{GB}^2\right)^2+16 N_{GB} \left(N_{GB}^2-1\right) \omega  (N_{GB} r-i \omega )+r \omega \right]\nonumber\\
&&\left.+r^5 f ^2 \left\{k^2 N_{GB}^2 \left(1-2 N_{GB}^2\right)^2 \left[8 \left(N_{GB}^2-N_{GB}^4\right) (\omega +i N_{GB} r)-5 i N_{GB} r\right]\right.\right.\nonumber\\
&&\left.\left.+4 \left(N_{GB}^2-N_{GB}^4\right) \omega ^2 \left[4 i N_{GB} \left(N_{GB}^2-N_{GB}^4\right) r+12 \left(N_{GB}^2-N_{GB}^4\right) \omega -5 i N_{GB} r\right]\right\}\right.\nonumber\\
&&\left.+4 \left(N_{GB}^2-N_{GB}^4\right) N_{GB}^3 r^3 f ^3 \left[k^2 \left(1-2 N_{GB}^2\right)^2 \left(2 N_{GB}^3 \omega -2 N_{GB} \omega +3 i r\right)\right.\right.\nonumber\\
&&\left.\left.-2 \left(N_{GB}^2-1\right) \omega ^2 \left(8 N_{GB}^3 \omega -8 N_{GB} \omega +11 i r\right)\right]+4 \left(N_{GB}^2-1\right)^2 N_{GB}^4 r f ^4 \left[-3 i k^2 N_{GB}^3 r\right.\right.\nonumber\\
&&\left.\left.+12 i \left(N_{GB}^2-N_{GB}^4\right) N_{GB} r \left(k^2 N_{GB}^2-3 \omega ^2\right)+8 \left(N_{GB}^2-1\right)^2 N_{GB}^4 \omega ^3\right]\right.\nonumber\\
&& +80 i \left(N_{GB}^2-1\right)^4 N_{GB}^9 \omega ^2 f ^5+2 r^9 \omega ^2 (\omega +2 i N_{GB} r),
\ea
\ba
i\alpha_3 B_3&=&
4 \left(N_{GB}^2-1\right)^2 N_{GB}^6 r^2 \omega  f ^3 \left[40 \left(N_{GB}^2-1\right) \omega ^2-3 k^2 \left(1-2 N_{GB}^2\right)^2\right]\nonumber\\
&&\left.+4 \left(N_{GB}^2-N_{GB}^4\right) N_{GB}^2 r^3 \omega  f ^2 \left[4 \left(2 N_{GB}^6-4 N_{GB}^4-5 N_{GB}^2+7\right) r \omega ^2\right.\right.\nonumber\\
&&\left.\left.+k^2 \left(1-2 N_{GB}^2\right)^2 \left(i N_{GB}^3 \omega -i N_{GB} \omega +4 r\right)\right]+i N_{GB}^2 r^5 f  \left[k^4 N_{GB} \left(1-2 N_{GB}^2\right)^4\right.\right.\nonumber\\
&&\left.\left.-k^2 \left(1-2 N_{GB}^2\right)^2 \omega  \left(4 N_{GB}^3 \omega -4 N_{GB} \omega -7 i r\right)+32 i \left(N_{GB}^6-2 N_{GB}^4+1\right) r \omega ^3\right]\right.\nonumber\\
&&\left.+80 \left(N_{GB}^2-1\right)^4 N_{GB}^8 \omega ^3 f ^4+r^7 \omega  \left[\left(-8 N_{GB}^4+8 N_{GB}^2+3\right) r \omega ^2\right.\right.\nonumber\\
&&\left.+k^2 N_{GB} \left(1-2 N_{GB}^2\right)^2 (4 N_{GB} r-i \omega )\right],
\ea
where
\ba
\alpha_3&=&N_{GB} r f  \left[2 \left(N_{GB}^2-1\right) N_{GB}^2 f +r^2\right]^2 \left\{N_{GB}^2 r^2 f  \left[4 \left(N_{GB}^2-1\right) \omega ^2-k^2 \left(1-2 N_{GB}^2\right)^2\right]\right.\nonumber\\
&&\left.+4 \left(N_{GB}^2-1\right)^2 N_{GB}^4 \omega ^2 f ^2+r^4 \omega ^2\right\}.
\ea

\subsection{Gauss-Bonnet corrections to metric perturbations in the sound channel}
\label{app:GB:sound}

The coefficients $A_4$ and $B_4$ in the equation for $Z_4$ \eq{Z4 eq} are given from
\ba
\alpha_4A_4&=&-4 N_{GB}^6 \left(N_{GB}^2-1\right)^2 r f ^4 \left\{k^2 N_{GB} \left[15 r-4 N_{GB} \left(N_{GB}^2-1\right) (3 N_{GB} r+i \omega )\right]\right.\nonumber\\
&&\left.+12 N_{GB} \left(N_{GB}^2-1\right) \omega ^2 \left[9 r-2 i N_{GB} \left(N_{GB}^2-1\right) \omega \right]\right\}-N_{GB} r^7 f  \left\{2 k^2 N_{GB} \left[N_{GB} \left(-8 N_{GB}^4\right.\right.\right.\nonumber\\
&&\left.\left.\left.+8 N_{GB}^2+1\right) r+i \left(20 N_{GB}^4-20 N_{GB}^2+1\right) \omega \right]+3 \omega ^2 \left[r+16 N_{GB} \left(N_{GB}^2-1\right) (N_{GB} r-i \omega )\right]\right\}\nonumber\\
&&+N_{GB} r^5 f ^2 \left\{k^2 N_{GB} \left[16 \left(N_{GB}^2-1\right)^2 N_{GB}^4 (12 N_{GB} r-7 i \omega )+4 i \left(N_{GB}^2-N_{GB}^4\right) (\omega +29 i N_{GB} r)\right.\right.\nonumber\\
&&\left.\left.+9 rN_{GB} \right]+12 \left(N_{GB}^2-N_{GB}^4\right) \omega ^2 \left[5 r+4 N_{GB} \left(N_{GB}^2-1\right) (N_{GB} r-3 i \omega )\right]\right\}\nonumber\\
&&+2 N_{GB}^2 \left(N_{GB}^2-N_{GB}^4\right) r^3 f ^3 \left\{12 N_{GB} \left(N_{GB}^2-1\right) \omega ^2 \left(-8 i N_{GB}^3 \omega +8 i N_{GB} \omega +11 r\right)\right.\nonumber\\
&&\left.+k^2 \left[36 \left(N_{GB}^2-N_{GB}^4\right) N_{GB} r+48 i \left(N_{GB}^2-1\right)^2 N_{GB}^4 \omega +4 i \left(N_{GB}^2-N_{GB}^4\right) \omega +13 N_{GB} r\right]\right\}\nonumber\\
&&+40 \left(N_{GB}^3-N_{GB}^5\right)^3 f ^5 \left[k^2+6 \left(N_{GB}^2-1\right) \omega ^2\right]+2 r^9 \left(2 k^2 N_{GB}^2-3 \omega ^2\right) (2 N_{GB} r-i \omega ),
\ea

\ba
\beta_4B_4&=&-16 \left(N_{GB}^3-N_{GB}^5\right)^3 f ^5 \left\{5 i \left(N_{GB}^2-1\right) N_{GB} r \omega  \left[k^2+6 \left(N_{GB}^2-1\right) \omega ^2\right]\right.\nonumber\\
&&\left.-6 k^2 \left(1-2 N_{GB}^2\right)^2 r^2-k^2 \left(N_{GB}^2-1\right) N_{GB}^2 \left[k^2+6 \left(N_{GB}^2-1\right) \omega ^2\right]\right\}\nonumber\\
&&+16 N_{GB}^7 \left(N_{GB}^2-1\right)^2 r^2 f ^4 \left\{k^4 N_{GB}^2 \left(6 N_{GB}^6-12 N_{GB}^4+5 N_{GB}^2+1\right)\right.\nonumber\\
&&\left.+k^2 \left[-3 N_{GB}^2 \left(N_{GB}^2-1\right)^2 \omega  (5 \omega +2 i N_{GB} r)+11 i N_{GB} \left(N_{GB}^2-1\right) r (\omega +4 i N_{GB} r)-11 r^2\right]\right.\nonumber\\
&&\left.+75 i N_{GB} \left(N_{GB}^2-1\right)^2 r \omega ^3\right\}-N_{GB}^2 r^8 f  \left\{k^4 N_{GB} \left[16 (N_{GB}-1) (N_{GB}+1) \left(8 N_{GB}^4\right.\right.\right.\nonumber\\
&&\left.\left.\left.-8 N_{GB}^2+1\right) N_{GB}^2+1\right]+k^2 \left[2 \left(N_{GB}^2-1\right) N_{GB} \left(112 N_{GB}^2 r^2+72 i N_{GB} r \omega -3 \omega ^2\right)\right.\right.\nonumber\\
&&\left.\left.+48 \left(N_{GB}^2-1\right)^2 N_{GB}^3 \left(8 N_{GB}^2 r^2-3 \omega ^2\right)+r (32 N_{GB} r+11 i \omega )\right]\right.\nonumber\\
&&\left.+6 i \left(24 N_{GB}^6-48 N_{GB}^4+5 N_{GB}^2+19\right) r \omega ^3\right\}-8 \left(N_{GB}^2-N_{GB}^4\right) r^4 f ^3 \left\{k^4 \left(N_{GB}^2-1\right) \right.\nonumber\\
&&\times\left.\left(24 N_{GB}^4-24 N_{GB}^2+1\right) N_{GB}^5+k^2 N_{GB}^2 \left[24 \left(N_{GB}^2-1\right)^3 N_{GB}^5 \omega ^2\right.\right.\nonumber\\
&&\left.\left.+\left(N_{GB}^2-N_{GB}^4\right) r (8 N_{GB} r-15 i \omega )+8 \left(N_{GB}^2-1\right)^2 N_{GB}^3 \left(12 N_{GB}^2 r^2-5 i N_{GB} r \omega -3 \omega ^2\right)\right.\right.\nonumber\\
&&\left.\left.-8 N_{GB} r^2\right]+24 i \left(N_{GB}^2-1\right)^2 \left(-N_{GB}^4+N_{GB}^2+6\right) N_{GB}^4 r \omega ^3\right\}\nonumber\\
&&+4 N_{GB}^3 r^6 f ^2 \left\{-k^4 N_{GB}^2 \left(N_{GB}^2-1\right) \left[12 \left(4 N_{GB}^8-8 N_{GB}^6+3 N_{GB}^4+N_{GB}^2\right)-1\right]\right.\nonumber\\
&&\left.+k^2 \left[4 \left(N_{GB}^2-1\right)^2 N_{GB}^2 \left(40 N_{GB}^2 r^2-27 i N_{GB} r \omega -3 \omega ^2\right)+\left(N_{GB}^2-1\right) N_{GB} r \right.\right.\nonumber\\
&&\left.\left.\times(56 N_{GB} r+i \omega )+24 \left(N_{GB}^2-1\right)^3 N_{GB}^4 \omega  (3 \omega -2 i N_{GB} r)+4 r^2\right]+12 i N_{GB} \left(N_{GB}^2-1\right)^2\right.\nonumber\\
&& \left.\left(-6 N_{GB}^4+6 N_{GB}^2+11\right) r \omega ^3\right\}-r^{10} \left\{2 k^4 \left(8 N_{GB}^7-8 N_{GB}^5+N_{GB}^3\right)\right.\nonumber\\
&&\left.+3 i \left(8 N_{GB}^4-8 N_{GB}^2-3\right) r \omega ^3-k^2 N_{GB} \left[16 N_{GB}^2 \left(1-2 N_{GB}^2\right)^2 r^2\right.\right.\nonumber\\
&&\left.\left.+2 i N_{GB} \left(24 N_{GB}^4-24 N_{GB}^2+1\right) r \omega +3 \left(8 N_{GB}^4-8 N_{GB}^2+1\right) \omega ^2\right]\right\},
\ea
where
\ba\label{alpha 4}
\alpha_4&=&N_{GB} r f  \left[2 \left(N_{GB}^2-1\right) N_{GB}^2 f +r^2\right]^2 \nonumber\\
&&\times\left\{-2 \left(N_{GB}^2-1\right) N_{GB}^4 f ^2 \left[k^2+6 \left(N_{GB}^2-1\right) \omega ^2\right]+N_{GB}^2 r^2 f  \left[k^2 \left(12 N_{GB}^4-12 N_{GB}^2+1\right)\right.\right.\nonumber\\
&&-\left.\left.12 \left(N_{GB}^2-1\right) \omega ^2\right]+r^4 \left(2 k^2 N_{GB}^2-3 \omega ^2\right)\right\},\\
\beta_4&=&\alpha_4\left[2 N_{GB}^2 (N_{GB}^2-1) r f +r^3\right].
\ea

\section{Pole-skipping point in the upper half plane of complex $\omega$}
\label{stringy GB upper}
\subsection{Stringy correction}

\label{stringy upper}

In the absence of the stringy correction, the point $r=r_0$ is a regular singularity of the second order differential equation \eq{Z eq form}. The special point is $k^2=\frac{3}{2}\omega^2$, at which the near horizon structure of the equation changes. As discussed in \cite{Blake:2019otz}, the pole-skipping point in the upper half plane can be identified via analysis at this special point.

When the stringy correction is taken into account, the above special point is expected to be shifted by a $\gamma$-correction of the general form
\be\label{k k1}
 k=\sqrt{\frac{3}{2}}\omega+\gamma k_1,
\ee
where $k_1$ is to be determined. The results of the pole-skipping points in the lower half plane in the main text are obtained under the implicit assumption that \eq{k k1} does not hold, even though $k_1$ is unknown.

Indeed, one cannot see $k_1$ directly from the expressions for $A$, $B$, $M_0$ and $M_1$ listed in Appendix \ref{app:stringy:sound}, as it is essentially a perturbative effect. However, further investigation reveals that $k_1$ can be extracted by considering the following requirement. Namely, {\it as a perturbative effect, the stringy correction should not change the fact that $r=r_0$ is a regular singularity}.
%As we shall see, this requirement will modify the location of the special point $k=\sqrt{3/2}\omega$ with an additional $\gamma$-dependent shift.

To extract $k_1$, one can first redefine the coefficients $A$ and $B$ by absorbing the $\gamma$-dependent terms, and rewrite the equation (of the same form as \eq{Z eq form}) for $Z_2$ in the sound channel as
\be\label{new eq Z2}
Z_2''+\tilde A Z_2'+\tilde B Z_2=0,
\ee
where $\tilde A=A-\gamma M_1$ and $\tilde B=B-\gamma M_0$. Then, the above requirement implies that  $\tilde A$ and $\tilde B$ should not have poles higher than $(r-r_0)^{-1}$  and $(r-r_0)^{-2}$, respectively. In particular,
\bi
    \item
    Inserting \eq{k k1} into $\tilde A$, there are two contributions from $A$ and $\gamma M_1$,
\ba\label{A tilde}
A&=&\frac{2 \sqrt{\frac{2}{3}} \gamma  k_1 r_0}{\omega  (r-r_0)^2}+\frac{-1-\frac{i \omega }{2 r_0}}{r-r_0}+\co\left[(r-r_0)^0\right],\nonumber\\
\gamma M_1&=&\gamma\left[\frac{\frac{32 \omega ^2}{r_0}+105 r_0}{(r-r_0)^2}-\frac{15 i \omega }{2 r_0 (r-r_0)}+\co\left[(r-r_0)^0\right]\right].
\ea
To ensure that the pole at $r=r_0$ of $\tilde A$  is still a regular singularity, the $(r-r_0)^{-2}$ terms must cancel, from which $k_1$ is determined as
\be\label{k1 solution}
k_1=\frac{1}{2} \sqrt{\frac{3}{2}} \omega  \left(\frac{32 \omega ^2}{r_0^2}+105\right).
\ee
    \item
Inserting \eq{k k1} into $\tilde B$ leads to two contributions
\ba\label{B tilde}
B&=&\frac{-\frac{\sqrt{\frac{2}{3}} \gamma  k_1 r_0}{\omega }-\frac{i \gamma  k_1}{\sqrt{6}}}{(r-r_0)^3}+\frac{\frac{5 \gamma  k_1}{\sqrt{6} \omega }-\frac{i \gamma  k_1}{2 \sqrt{6} r_0}+\frac{i \omega }{2 r_0}+1}{(r-r_0)^2}+\co\left[(r-r_0)^{-1}\right],\nonumber\\
\gamma M_0&=&\gamma\left[\frac{-\frac{8 i \omega ^3}{r_0^2}-\frac{16 \omega ^2}{r_0}-\frac{105 r_0}{2}-\frac{105 i \omega }{4}}{(r-r_0)^3}+\frac{-\frac{4 i \omega ^3}{r_0^3}+\frac{40 \omega ^2}{r_0^2}-\frac{45 i \omega }{8 r_0}+\frac{525}{4}}{(r-r_0)^2}+\co\left[(r-r_0)^{-1}\right]\right].\nonumber\\
\ea
The requirement of being a regular singularity at $r=r_0$ now means the $(r-r_0)^{-3}$ terms must cancel. This is trivially satisfied by inserting the expression \eq{k1 solution} for $k_1$.
\ei

Now from \eq{A tilde} and \eq{B tilde}, we see that at the shifted special point \eq{k k1} with $k_1$ given by \eq{k1 solution}, the pole structures of $\tilde A$ and $\tilde B$ are
\ba
\tilde A&=&\frac{\tilde A_{-1}}{r-r_0}+\co\left[(r-r_0)^0\right],\\
\tilde B&=&\frac{\tilde B_{-2}}{(r-r_0)^2}+\left[(r-r_0)^{-1}\right],
\ea
where
\ba
\tilde A_{-1}&=&-1-\frac{i \omega }{2 r_0}+\frac{15 i \gamma  \omega }{2 r_0},\\
\tilde B_{-2}&=&1+\frac{i (1-15 \gamma ) \omega }{2 r_0}.
\ea
For a series solution
\be
Z_2=(r-r_0)^\rho\sum_{n=0}Z_{2n}(r-r_0)^n,
\ee
equation \eq{new eq Z2} leads to the indicial equation
\be
\rho(\rho-1)+\rho \tilde A_{-1}+\tilde B_{-2}=0,
\ee
which gives two solutions
\be
\rho_1=1,\ \ \ \rho_2=1+\frac{i\omega}{2\pi T}.
\ee
So, solutions regular at the horizon with $\rho=0,1,2$ are given by $\omega=i2\pi T,0,-i 2\pi T$,   respectively. Note that in this case, $\omega=-i2\pi T$ does not give pole-skipping, in that $G^R_{T^{00}T^{00}}$ is independent of $\delta \omega/\delta k$, the parameter characterizing the way the point is approached . The $\omega=0$ case corresponds to the hydrodynamic mode, which is already characterized by the pole in the two point function, and therefore should not concern us. One can show that $\omega=+i2\pi T\equiv\omega_*$  is indeed the desired pole-skipping point (dependent on $\delta\omega/\delta k$) in the upper plane. Inserting this value into \eq{k1 solution} leads to
\be\label{app:stringy sound upper k1}
k_1=-i\sqrt{6}\frac{23}{2}r_0,
\ee
and the butterfly velocity
\be
v_B=\frac{\omega_*}{\sqrt{3/2}\omega_*+\gamma k_1}=\sqrt{\frac{2}{3}}\left(1+\frac{23}{2}\gamma\right),
\ee
which agrees with the result obtained by analyzing the $vv$ component of Einstein equation in \cite{Grozdanov:2018kkt}.
\subsection{Gauss-Bonnet correction}
\label{GB upper}

For metric perturbations in the sound channel in the GB corrected background\footnote{ Similar discussion is also given in \cite{overlapping}.}, the near horizon structure of the differential equation \eq{Z4 eq} depends on whether $k^2=3\omega^2/(2N^2_{GB})$, as can be easily seen from the results in Appendix \ref{app:GB:sound}, and in particular, the expression for $\alpha_4$ in \eq{alpha 4}.

The results obtained in the main text are implicitly under the assumption that $k^2\neq 3\omega^2/(2N^2_{GB})$.  On the special point $k^2= 3\omega^2/(2N^2_{GB})$, however, the coefficients in \eq{Z4 eq} have different singular structures
\ba
A&=&\frac{A_{-1}}{r-r_0}+\co\left[(r-r_0)^0\right],\\
B&=&\frac{B_{-2}}{(r-r_0)^2}+ \co\left[(r-r_0)^{-1}\right],
\ea
where
\ba
A_{-1}&=&-1-\frac{i \omega }{2 N_{GB} r_0},\\
B_{-2}&=&{1+\frac{i \omega }{2 N_{GB} r_0}}.
\ea
 Then the indicial equation gives
\be
\rho_1=1,\ \ \ \rho_2=1+\frac{i \omega }{2 N_{GB} r_0},
\ee
from which the argument in the previous subsection immediately leads to the conclusion that $\omega=+i2\pi T$  is the desired pole-skipping point in the upper half plane. Inserting this value into $k^2= 3\omega^2/(2N^2_{GB})$ leads to  
\be\label{app:GB upper kstar}
k_*=i\sqrt{6}r_0=i\sqrt{6}\pi \frac{T}{N_{GB}},
\ee
and the butterfly velocity $v_B=\sqrt{2/3}N_{GB}$, which agrees with the result of \cite{Grozdanov:2018kkt}.

\section{Corrections to $k_1$ in three channels of metric perturbations}
\label{app:corrections to k1}
 {
For the stringy correction
\bi
    \item
    sound channel: $\omega_*=+i2\pi T$, $k_*=i\sqrt{6}r_0-\gamma i\sqrt{6}\frac{23}{2}r_0$ (from \eq{app:stringy sound upper k1});
    \item
    shear channel: $\omega_* =-i2\pi T$, $k_*=k_1=\sqrt{6}r_0-\gamma737 \sqrt{\frac{3}{2}}  r_0$ (from \eq{stringy shear k1 sol});
    \item
    scalar channel: $\omega_*=-i2\pi T$, $k_*=k_1=i\sqrt{6}r_0-\gamma 473 i \sqrt{\frac{3}{2}}r_0$.
\ei
For the Gauss-Bonnet correction
\bi
 \item
    sound channel: $\omega_*=+i2\pi T$, $k_*=i\sqrt{6}r_0$ (from \eq{app:GB upper kstar});
    \item
    shear channel: $\omega_* =-i2\pi T$, $k_*=k_1=\sqrt{\frac{2r_0^2 \left(3+8 N_{GB}^2-8 N_{GB}^4\right)}{\left(1-2 N_{GB}^2\right)^2}}$ (from \eq{GB shear k1 sol});
    \item
    scalar channel: $\omega_*=-i2\pi T$, $k_*=k_1=\sqrt{ \frac{2 r_0^2\left(3+8 N_{GB}^2-8 N_{GB}^4\right) }{8 N_{GB}^4-8 N_{GB}^2-1}}$.
\ei
All the expressions for $k_*$ at $\gamma=0$ or $N_{GB}=1$ recover the uncorrected results.
}

\bibliographystyle{JHEP}
\bibliography{reference}
\end{document}